\documentclass[sigconf]{acmart}

\AtBeginDocument{%
  \providecommand\BibTeX{{%
    \normalfont B\kern-0.5em{\scshape i\kern-0.25em b}\kern-0.8em\TeX}}}

\copyrightyear{2025}
\acmYear{2025}
\setcopyright{acmlicensed}\acmConference[UIST '25]{The 38th Annual ACM Symposium on User Interface Software and Technology}{September 28-October 1, 2025}{Busan, Republic of Korea}
\acmBooktitle{The 38th Annual ACM Symposium on User Interface Software and Technology (UIST '25), September 28-October 1, 2025, Busan, Republic of Korea}
\acmDOI{10.1145/3746059.3747790}
\acmISBN{979-8-4007-2037-6/2025/09}

\usepackage{siunitx}
\usepackage{multirow} 
\usepackage{enumitem}

\begin{document}
\newcommand{\mynote}[1]{\todo[size=\footnotesize, color=blue!20]{#1}}
\newcommand{\ludwignote}[1]{\todo[size=\footnotesize, color=red!20]{#1}}

\newcommand{\system}[0]{Adaptique}

\title[\system{}]{\system{}: Multi-objective and Context-aware Online Adaptation of Selection Techniques in Virtual Reality}


\author{Chao-Jung Lai}
\affiliation{%
  \department{Department of Computer Science}
  \institution{University of Toronto}
  \city{Toronto}
  \state{Ontario}
  \country{Canada}
}
\affiliation{%
  \department{Department of Computer Science}
  \institution{University of California, San Diego}
  \city{La Jolla}
  \state{California}
  \country{USA}
}
\email{chaojunglai@cs.toronto.edu}

\author{Mauricio Sousa}
\affiliation{%
  \department{Department of Computer Science}
  \institution{University of Toronto}
  \city{Toronto}
  \state{Ontario}
  \country{Canada}
}
\email{mauricio@dgp.toronto.edu}

\author{Tianyu Zhang}
\affiliation{%
  \department{Department of Computer Science}
  \institution{University of Toronto}
  \city{Toronto}
  \state{Ontario}
  \country{Canada}
}
\affiliation{%
  \department{Department of Computer Science}
  \institution{University of Rochester}
  \city{Rochester}
  \state{New York}
  \country{USA}
}
\email{tianyuz@cs.toronto.edu}

\author{Ludwig Sidenmark}
\affiliation{%
  \department{Department of Computer Science}
  \institution{University of Toronto}
  \city{Toronto}
  \state{Ontario}
  \country{Canada}
}
\email{lsidenmark@dgp.toronto.edu}

\author{Tovi Grossman}
\affiliation{%
  \department{Department of Computer Science}
  \institution{University of Toronto}
  \city{Toronto}
  \state{Ontario}
  \country{Canada}
}
\email{tovi@dgp.toronto.edu}

\renewcommand{\shortauthors}{Lai \emph{et al.}}


\begin{abstract}

Selection is a fundamental task that is challenging in virtual reality due to issues such as distant and small targets, occlusion, and target-dense environments. Previous research has tackled these challenges through various selection techniques, but complicates selection and can be seen as tedious outside of their designed use case. We present \emph{\system}, an adaptive model that infers and switches to the most optimal selection technique based on user and environmental information. \system{} considers contextual information such as target size, distance, occlusion, and user posture combined with four objectives: speed, accuracy, comfort, and familiarity which are based on fundamental predictive models of human movement for technique selection. This enables \system{} to select simple techniques when they are sufficiently efficient and more advanced techniques when necessary. We show that \system{} is more preferred and performant than single techniques in a user study, and demonstrate \system{}'s versatility in an application.



\end{abstract}



\begin{teaserfigure}
  \centering
  \includegraphics[width=.85\textwidth]{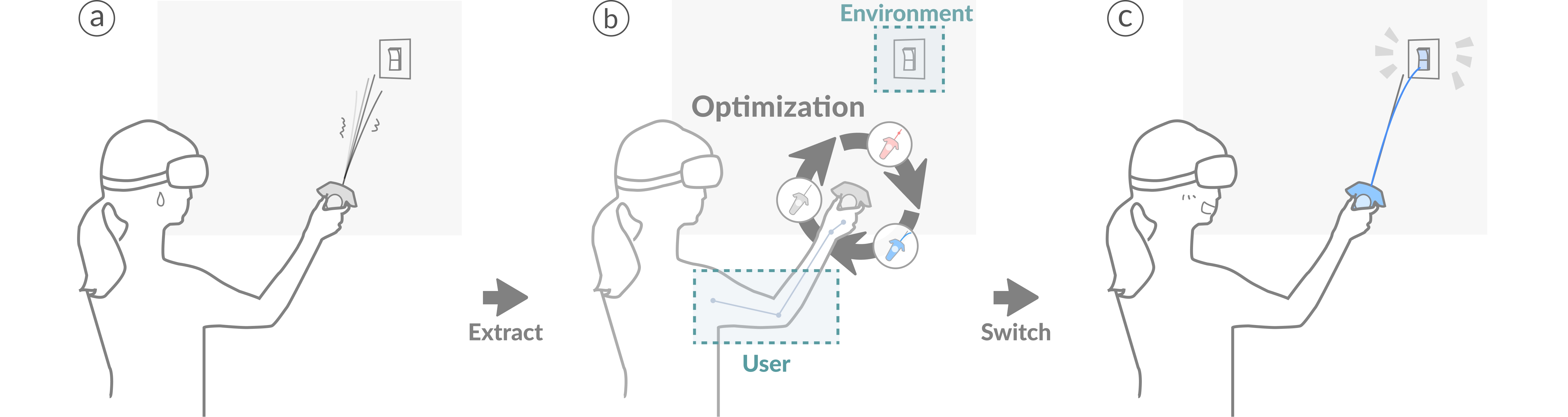}
  \caption{\system{} switches the selection technique based on environmental and user-based factors, and considers multiple objectives for VR selection. In this example, the user attempts to select the light switch on the far wall to light up the room. (a) Since the light switch is small and far, the user has difficulty selecting it with normal RayCasting. (b) \system{} continuously senses the environment and user state to find the most optimal selection technique for current use. (c) \system{} switches the selection technique to StickyRay, snapping the ray to the nearest target to assist the user in accurately and comfortably selecting the light switch.}
  \Description{This image is a three-part illustration (labeled a, b, and c) demonstrating the process of interacting with a virtual environment with Adaptique. 
  Figure (a) shows a person wearing a VR headset and holding a controller, using a laser pointer (representing RayCasting) aimed at a distant, small virtual light switch. The user appears to be sweating, indicating effort or difficulty. A right-facing arrow labeled Extract points to the next step (b), representing the extraction of contextual information. 
  Figure (b) illustrates an Optimization process where both the user and the environment are highlighted as inputs. Arrows loop between three selection techniques—StickyRay (blue), RayCasting (gray), and RayCursor (red)—symbolizing the iterative process of determining the most effective method for the task. 
  Figure (c) shows the user successfully interacting with the virtual light switch using a blue laser (representing StickyRay). The light switch is illuminated, indicating that the selection task is completed. This step is labeled Switch, signifying the transition to the StickyRay technique as the final selection method.}
  \label{fig:teaser}
\end{teaserfigure}


\maketitle

\section{Introduction}
Selection tasks in extended reality (XR) can be challenging in dynamic environments due to factors such as small or distant targets and occlusion~\cite{argelaguet2013Survey}. Furthermore, XR environments often change rapidly, with virtual contents changing and users moving or altering their attention within the 3D space~\cite{lindlbauer2022Future}. For example, a user might be selecting buttons on a large panel, which requires only a simple and easy selection technique. Then, they might shift to examining components within a complex 3D assembly file they just opened. Since these components are small and cluttered, the user needs a precise selection technique designed to target fine details in a dense environment. Later, they might interact with an IoT light switch on a distant wall to view a real-world object clearly. Because the switch is small and located far away from their reach, they need a technique that can effectively handle distant and small objects. These scenarios represent three distinct environments, making the use of a single technique for all tasks difficult. 

Previous works have addressed some of the selection challenges in XR, such as selecting small, distant, and occluded objects, or selection in a dense environment~\cite{steinicke2006Object, yu2020FullyOccluded} by commonly manipulating the amplitude and width of targets to make pointing and selection easier. However, these techniques are often tailored to specific scenarios and become overly complicated and cumbersome when applied outside their intended context. Manually switching between techniques adds extra workload to the user, who must identify the current context and needs and then perform interaction to switch the technique. To alleviate the burden on users, researchers have proposed adaptive systems that change the current selection technique based on the environment~\cite{cashion2013Optimal, lacoche2019Machine} or user factors~\cite{sidenmark2022Weighted}. However, these works only focus on a single objective for switching or require bespoke scoring algorithms that do not consider the different selection trade-offs (e.g., speed versus accuracy).

We propose \emph{\system{}}, a novel multi-objective adaptive virtual reality (VR) optimization system that selects the real-time optimal pointing-based selection technique (\autoref{fig:teaser}) based on environmental and user factors. In contrast to previous work, \system{} consists of multiple objectives based on established pointing-based performance metrics such as speed, accuracy, comfort, and familiarity and their associated established models. This enables \system{} to consider the different trade-offs for selection to holistically select the most optimal technique. Furthermore, using established performance models as optimization objectives ensures reliability and compatibility across different techniques as opposed to bespoke scoring algorithms or machine learning models trained on small data sets. The developer only has to define how a technique affects the model inputs for a given target (i.e., target width and amplitude).

To select the most optimal technique, \system{} senses the contextual information of selectable targets and the user (\autoref{fig:teaser}b) and sends the information to candidate techniques that apply their selection mechanism, which adjust the effective target width and amplitude. The adjusted values are then used as input for the optimization objectives, which calculate the current most effective selection technique for all selectable targets. \system{} will then make a decision in real-time and switch the technique when the performance reaches the predefined threshold of improvement (\autoref{fig:teaser}c). In our current implementation, we included a set of established pointing-based selection techniques of normal RayCasting, StickyRay~\cite{grossman2005Bubble, lu2020Investigating}, and RayCursor~\cite{baloup2019RayCursor} as candidate selection techniques. These techniques employ distinct mechanisms to modify the effective target size and amplitude compatible with \system{} and together cover common scenarios of normal selection, small targets, dense environments, and target occlusion.

We showcased \system{}'s utility and applicability in a VR indoor application where \system{} smoothly switches the selection tool to a more suitable one when the content of the user's interest changes and the task becomes hard with the current tool. Furthermore, our user study highlighted the importance of adaptivity, as using the same technique in different scenarios can lead to difficulty and negatively impact the user experience. We show how \system{} outperformed the use of singular techniques in selection time, movement, and error rate, and was also preferred by the majority of study participants. In sum, the contributions of this work are:

\begin{itemize}
    \item \system{}, a real-time multi-objective adaptive optimization system for selection techniques in VR.
    \item An application with various selection tasks that showcases the versatility and utility of \system{} in various natural selection contexts.
    \item The results of a user study that demonstrate \system{} benefits, and show how \system{} outperformed singular techniques in multiple performance metrics and user preference when used across various environments for selection.
\end{itemize}
\section{Related Work} \label{sec:related-work}
\system{} builds on common selection challenges in XR, the techniques designed to address these challenges, human selection performance models, and context-aware adaptive systems.


\subsection{XR Selection Techniques} \label{sec:related-work:xr-interaction-techniques}
RayCasting is one of the common selection techniques in the XR due to its ability to select targets beyond the user's reach by pointing with a ray extending from the user's hand or controller~\cite{mine1995virtual, laviola20173D}. However, selecting a small or distant object requires higher accuracy because of its small visible area and the tremor of the hand amplified along the ray. In addition, in dense environments, targets may be occluded, resulting in the requirement of physically changing the point of view to be able to see the target. Dense environments also increase the chance of erroneous selection due to the proximity of targets to one another. To address these challenges, various interaction techniques have been proposed.

To enhance the selection of small targets in the 3D space, researchers have proposed techniques that dynamically enlarge the size of objects to expand the interactable area~\cite{argelaguet2008Improving}, progressive refinement techniques that require steps following the initial action to improve precision~\cite{kopper2011Rapid, grossman2006Design, molina2023Twostep, kyto2018Pinpointing}, snapping mechanisms to decrease the precision requirement by enlarging the effective size~\cite{steinicke2006Object, gabel2023Redirecting, lu2020Investigating}.
Other works have tackled the problem of selection in dense environments. In addition to the previously mentioned progressive refinement techniques that also help disambiguation in dense environments, some approaches use extra degrees of freedom to specify the depth of the target. For example, Depth Ray~\cite{grossman2006Design}, RayCursor~\cite{baloup2019RayCursor}, ClockRay~\cite{wu2023ClockRay}, and Alpha Cursor~\cite{yu2020FullyOccluded} utilize an extra cursor along the ray that is controlled by the forward-backward movement of the hand, swiping on the trackpad, or wrist rotation. MultiFingerBubble~\cite{delamare2022MultiFingerBubble} uses multiple rays of individual fingers to select between nearby objects by flexing the corresponding finger. Other methods use visual aids, such as mirrors, that display occluded objects from different perspectives and make them visible~\cite{lee20203D}. 

Although these techniques make the selection task easier in their designed cases, 
their standalone use can be overly complex or inefficient when applied outside of their intended context. Adaptique builds on their individual strengths by embedding them within a context-aware adaptive system, allowing each technique to be leveraged where it works best
in dynamic XR applications.

\subsection{Human Performance Models on Selection Tasks} \label{sec:related-work:point-selection-models}
Researchers have developed various models to evaluate and predict user performance of pointing selection tasks, focusing on speed, accuracy, and comfort. We evaluate these factors using Fitts' Law~\cite{fitts1954Information} for speed, the end-point distribution model~\cite{yu2019Modeling} for accuracy, and the Consumed Endurance model~\cite{hincapie-ramos2014Consumed} for comfort. 


In Fitts' law, 
the predicted time needed to select a target based on the target's distance and width is formulated as $MT = a + b \cdot log_2(\frac{2A}{W})$~\cite{fitts1954Information}. Here, $A$ represents the amplitude of movement to the target, $W$ is the width of the target along the axis of motion, and the constants $a$ and $b$ are determined by empirical linear regression. The logarithmic term is the index of difficulty ($ID$) of the task. Though originally applied to 1D selection tasks, it has shown good applicability higher dimension spaces.
For example, Shannon formulation defined the movement time as $MT = a + b \cdot log_2(\frac{A}{W}+1)$. To capture the target geometry, W'-model adjusts the definition of 1D width as the cross-section width along the direction of cursor movement~\cite{mackenzie1992Extending}. We adopted this model due to its simplicity, its ability to deal with non-rectangular geometries, and its good fitting result in 2D tasks.
In the virtual environment, the Shannon formulation has been used in raycasting tasks because raycasting does not require z-axis movements~\cite{gabel2023Redirecting} with two rotation axes as its dominant degree of freedom (DoF)~\cite{argelaguet2013Survey}. The target width and amplitude are represented in angular size form to consider the depth~\cite{yu2019Modeling}. For interaction that requires a higher degree of freedom in translation, such as virtual hand pointing, the 3D Fitts' Law is used more frequently~\cite{clark2020Extending, saffo2021Remote}.

Endpoint distribution models describe selection behaviors by analyzing the spatial distribution of endpoints during pointing tasks. In XR, models such as the EDModel explore how different factors such as target size, target shape, movement amplitude, and target depth affect the distribution characteristic based on a bivariate Gaussian distribution~\cite{yu2019Modeling}. Combined with Bi's method~\cite{bi2016Predicting}, this model can estimate selection accuracy by integrating the probability density function for the target region into its control space. 

In addition to time and accuracy, user performance is influenced by physical factors such as fatigue and overall comfort. For example, the gorilla arm effect occurs when people feel fatigued in their arms and shoulders after performing mid-air interactions for a long time. Models such as Consumed Endurance (CE)~\cite{hincapie-ramos2014Consumed} and RULA~\cite{mcatamney1993RULA} characterize this ergonomic factor from a biomechanics perspective, relying on physical data such as user postures, arm weights, muscle endurance, and other relevant information.
In our work, we utilized these models to evaluate the most suitable selection techniques for the context regarding speed, accuracy, and comfort.

\subsection{Adaptive Systems for Interaction Techniques} \label{sec:adaptive-system}

Recent studies have increasingly highlighted the importance of context-aware adaptive systems in XR interactions~\cite{grubert2017Pervasive, evangelistabelo2022AUIT, lindlbauer2019ContextAware, johns2023Pareto, tahara2020Retargetable, cheng2021SemanticAdapt, evangelistabelo2021XRgonomics, cheng2023InteractionAdapt, gebhardt2019Learning}, especially in mixed reality environments due to their connection to the dynamic physical world. These works have, for example, adapted the layout of virtual content for various kinds of factors, such as the relationship between virtual and physical objects~\cite{cheng2021SemanticAdapt, tahara2020Retargetable}, ergonomics~\cite{evangelistabelo2021XRgonomics, johns2023Pareto}, physical space~\cite{cheng2023InteractionAdapt}, or user's intention~\cite{gebhardt2019Learning}. These systems are typically implemented through combinatorial optimization, rule-based systems, or data-driven methods such as reinforcement learning.
Most XR adaptation systems focus on adapting the layout of virtual content~\cite{cheng2021SemanticAdapt, cheng2023InteractionAdapt, evangelistabelo2021XRgonomics, evangelistabelo2022AUIT, lindlbauer2019ContextAware}, where elements can be freely moved and placed. In contrast, our work assumes that the content and interactable targets in XR are relatively static, as they may represent physical objects that users want to keep intact, with minimal changes to the surrounding environment. Instead, we believe that the user and their interaction should adapt to the current context. Inspired by layout adaptation systems, we employed a multi-objective optimization framework for adaptation due to its simplicity, scalability, and ability to balance multiple objectives in a controllable and interpretable way~\cite{oulasvirta2020Combinatorial}.


There has been limited exploration into adapting selection techniques or selecting appropriate input tools in XR. Although some early efforts have explored the adaptation of selection techniques in virtual environments, 
these studies either rely on rule-based switching tied to fixed condition, focus on user preference modeling, or remain conceptual without an implemented system~\cite{octavia2011Adaptation, cashion2013Optimal, lacoche2019Machine}. As such, they do not support generalizable, real-time switching based on quantified trade-offs across multiple performance objectives.
Other works have adapted the modality of selection techniques based on availability or stability. For example,~\citet{sidenmark2022Weighted} switched from gaze input to a controller or head-based input when the gaze signal quality dropped, and~\citet{yigitbas2019Contextaware} switched to gaze input when controllers were unavailable. More recently, selection techniques have been included as part of XR layout adaptation systems~\cite{cheng2023InteractionAdapt}. In contrast, \system{} builds on these works in multiple ways: (1) we consider multiple adaptation objectives to reflect selection trade-offs; (2) we use established performance models as optimization objectives to simplify comparability between techniques; and (3) we integrate these models with real-time environmental sensing of selectable candidates and their relationships with each other for online adaptation.

\section{\system{}}
\label{sec:system}

\begin{figure*}[t]
    \centering
    \includegraphics[width=.9\linewidth]{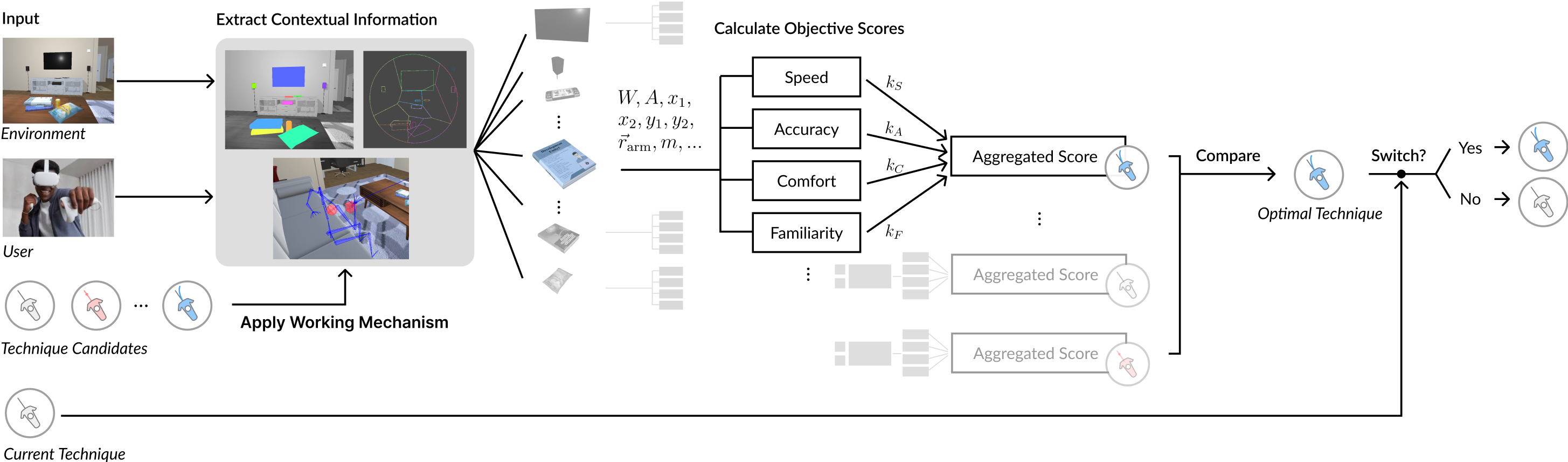}
    \caption{The \system{} pipeline extracts user and environmental input, applies the working mechanism of selection techniques to calculate model input for each selectable object and technique, aggregates the objects' objective scores for each technique, finds the optimal technique for the interactable objects, and switches the technique if the performance gain is above a threshold.}
    \Description{This image illustrates the Adaptique pipeline, which is designed to optimize the selection technique based on user and environmental input. The pipeline is organized into several stages: 
    On the left, two input sources are depicted: the Environment and the User. The environment is represented by a virtual living room setup, while the user is shown interacting within a VR space. The current selection technique and technique candidates (RayCasting, StickyRay, RayCursor, and others) are also presented. 
    The next step, labeled Extract Contextual Information, demonstrates how the system gathers spatial data from the environment, including object positions and user interactions, as well as user data, visualized as a skeletal representation of the user’s posture. 
    At this stage, the working mechanisms of different selection techniques are applied to the extracted data. In the central portion of the figure, the system calculates objective scores for each technique and target. Multiple objectives are depicted, including Speed, Accuracy, Comfort, Familiarity, and others. Each score is assigned a specific weighting factor, and these scores are aggregated into a single objective score for each technique. 
    To the right, the aggregated scores for each technique are compared. The pipeline identifies the Optimal Technique and evaluates whether switching to a different technique will result in a performance improvement. If the performance gain surpasses a predetermined threshold, the system switches techniques; otherwise, it maintains the current technique. The flowchart concludes with the final technique selection, with branches labeled Yes or No to indicate whether a switch has been made.
    }
    \label{fig:pipeline}
\end{figure*} 

We define the problem of adapting the interaction technique as follows: given a virtual environment with all inferred selection targets, the system will choose the interaction technique that maximizes the performance of the selection task in terms of four objectives: speed, accuracy, comfort, and familiarity. Speed, accuracy, and comfort are three common metrics used to evaluate interaction technique performance~\cite{Bergstrom2021Study}. However, while advanced interaction techniques improve performance in selection tasks, they often come with trade-offs such as increased complexity in control and limited interaction expressiveness (e.g., lack of continuous interaction). Overcoming these drawbacks necessitates user familiarity~\cite{scarr2011Dips}. 
Therefore, we consider it as one of the inputs in our system. We quantify these metrics and give objective scores to aid in our optimization process. The system would then post-process the data and switch the interaction technique for the user. 
As illustrated in \autoref{fig:pipeline}, the system processes each frame by:
\begin{enumerate}[topsep=1pt]
    \item Acquire the targets within the interaction space.
    \item Extract contextual information from the user and targets.
    \item Apply the working mechanism of each candidate technique on all selectable targets to use as model objective input.
    \item Calculate and aggregate the objectives for each candidate technique.
    \item Activate a switch if a more optimal one shows a consistent improvement in overall performance.
\end{enumerate}
Sections~\ref{sec:context-information}--\ref{sec:adaptation-and-switching} detail each pipeline component, while Sections~\ref{sec:technique}--\ref{sec:implementation} explain how specific techniques are integrated and describe system implementation.
\subsection{Extracting Contextual Information} \label{sec:context-information}
We first extract contextual information from the environment and the user that will be used as input for adaptation. Pointing-based selection techniques in XR can be broadly categorized into ray-based techniques, which are effectively 2D techniques in the pointing space, and techniques that use a 3D point, thus incorporating depth information to infer the pointed target~\cite{argelaguet2013Survey}. To support both types of techniques, we extract both the 2D and 3D information of selectable targets. We assume that the adaptation should be based only on the user's temporal area of attention, ensuring that only interaction-relevant targets are considered for adaptation and reducing unnecessary computation. Accordingly, we define the interaction space as a cone originating from the pointing direction, with a radius forming an angle of $r_c$ degrees~\cite{sidenmark2022Weighted}. All objects within this area are included as input for adaptation.

After defining the set of objects for interaction, we extract the most fundamental contextual information to serve as the basis on which techniques apply their working mechanisms to calculate model objective input. This includes the positions and sizes of all targets in both 3D and 2D space relative to the controller. For 3D information, we provide the 3D target positions, shapes, and sizes relative to the controller position. For 2D information, we project all targets' 3D meshes onto a plane perpendicular to the controller's pointing direction. We adopt controller-based projection, as occlusion relative to the controller is more relevant for determining technique applicability, and targets visible to the eyes may still be unreachable given the controller's pose. The projected targets are scaled to ensure that their visual size remains consistent. To incorporate occlusion, we calculate a convex mesh of the projected vertices to form an outline polygon using CGAL~\cite{cgal2024}. The outline polygon is then clipped by other overlapping polygons that occlude the object using Clipper2~\cite{clipper2024}. Finally, we recalculate the target centroid ($c$) using the final clipped polygon. For each target, we provide the final 2D outline polygon that defines its activation region, along with its position relative to the controller position.

In addition, we incorporate user-side information, including the user's posture and their current selection technique, as the adaptation input. Since most VR systems provide tracking only via controllers and the head-mounted display (HMD), we estimate the user's current posture using inverse kinematics based on the HMD and controller position. Both the 2D and 3D environmental information are updated every frame along with the user information.
\subsection{Objectives}\label{sec:objectives}
To comprehensively consider and balance different interaction goals, \system{} leverages multiple objectives to find the overall best technique given the contextual information provided. Each objective defines a set of parameters that each interaction technique has to provide for every interactable object to calculate objective scores. Then, all objective scores are calculated for each target and interaction technique. The objective scores of the techniques are then aggregated to a final overall score for each technique. 

In our implementation of \system{}, we use common performance metrics and formalize them with established models of human performance and movement. The system can be expanded to include more objectives and individual objectives can be altered or replaced to better suit the chosen interaction techniques.

\subsubsection{Speed} \label{sec:objective-speed}

Speed ($S_S$) is one of the most common performance metrics and selection objectives. We define speed score using the widely adopted index of difficulty (ID) formula in the Shannon formulation of Fitts' Law 
\begin{equation}
    S_S = -log_2(\frac{A}{W} + 1),
    \label{eq:shannon}
\end{equation}
which states that the difficulty of selection, and thus the expected selection time, increases with larger movement amplitude and smaller target width~\cite{mackenzie1992Extending}. The formula relies on two main parameters, movement amplitude $A$, and the target size $W$, both defined in visual angular space(\autoref{fig:scoring}a). For this work, we define $W$ as the effective width of the target activation region along the pointing path. The pointing path is defined as the angular trajectory from the current pointing direction to the target centroid, which has been shown to play a greater role than its visual boundaries on selection time~\cite{vanacken2007Exploring, grossman2007Modeling}. The movement amplitude $A$ is the angular distance the ray needs to travel along that path to the aiming center of the target, which we define as the centroid of the activation region. To ensures that higher performance yields higher scores, we add a negative sign to the objective. Finally, note that Fitts' Law commonly requires the fitting of additional parameters ($a$ and $b$) based on performance data to predict the selection speed. As all techniques in our implementation are based on controller pointing, we assume that these remain consistent between techniques, thus removing the need for fitting additional parameters and collecting user data. These should be added if \system{} is expanded to multiple pointing modalities.

\subsubsection{Accuracy}
Accuracy ($S_A$) is another critical metric that reflects the reliability of selection. We adopt the \emph{EDModel} by \citet{yu2019Modeling}. The EDModel defines an endpoint distribution for pointing-based selection. To calculate the probability of successful selection, we integrate the distribution with the target activation boundary, which we define as our accuracy score:

\begin{equation}
    S_A=\iint_D \frac{1}{2 \pi \sigma_x \sigma_y} \exp \left(-\frac{\left(x-\mu_x\right)^2}{2 \sigma_x^2}-\frac{y^2}{2 \sigma_y^2}\right) dx dy.
    \label{eq:endpoint}
\end{equation}

The variables $\mu_x$, $\sigma_x$, and $\sigma_y$ are derived via regression using empirical endpoint data~\cite{yu2019Modeling}. In the integral of \autoref{eq:endpoint}, we define the $x$-axis as the direction of movement, the $y$-axis as perpendicular to the direction of movement, and $D$ as the target activation region. We use the same definition for the direction of movement as in our speed objective. To simplify the integration, we approximate $D$ with a rectangle bounded by coordinates $(x_1$, $x_2$, $y_1$, $y_2)$--the intersection points of the activation region boundary with the axes  (\autoref{fig:scoring}b). To derive a closed-form approximation, we leverage the integral identity of the Gaussian distribution over a bounded interval, which can be expressed using the error function (erf). The error function is defined as $\text{erf}(z) = \frac{2}{\sqrt{\pi}} \int_0^z e^{-t^2} dt$, representing the cumulative distribution of a standard normal random variable. By substituting the limits of integration corresponding to the boundaries of the activation region $D$, the double integral in \autoref{eq:endpoint} is simplified into a product of one-dimensional error function terms for both the $x$-axis and $y$-axis, expressed as:

\begin{equation}
S_A = \frac{1}{4} \left( \text{erf} \left( \frac{y_1}{\sqrt{2}\sigma_y} \right) - \text{erf} \left( \frac{y_2}{\sqrt{2}\sigma_y} \right) \right) \left( \text{erf} \left( \frac{\mu - x_2}{\sqrt{2}\sigma_x} \right) - \text{erf} \left( \frac{\mu - x_1}{\sqrt{2}\sigma_x} \right) \right).
    \label{eq:endpoint-simplified}
\end{equation}

\subsubsection{Comfort}

Comfort plays a vital role in the usability of VR, as physical fatigue, commonly referred to as the gorilla-arm effect~\cite{boring2009scroll}, can significantly affect user performance. To define comfort ($S_C$), we adopt a modified Strength metric from the Consumed Endurance (CE) model to quantify arm fatigue caused by selection~\cite{hincapie-ramos2014Consumed}. In the CE model, the torque exerted on the shoulder must counteract gravitational torque $\vec{g}$:
\begin{equation}
    \vec{T}_{shoulder} = || \vec{r}_\text{arm} \times m\vec{g}||.
    \label{eq:shoulder-torque}
\end{equation}

\begin{figure}
    \centering
    \includegraphics[width=1\linewidth]{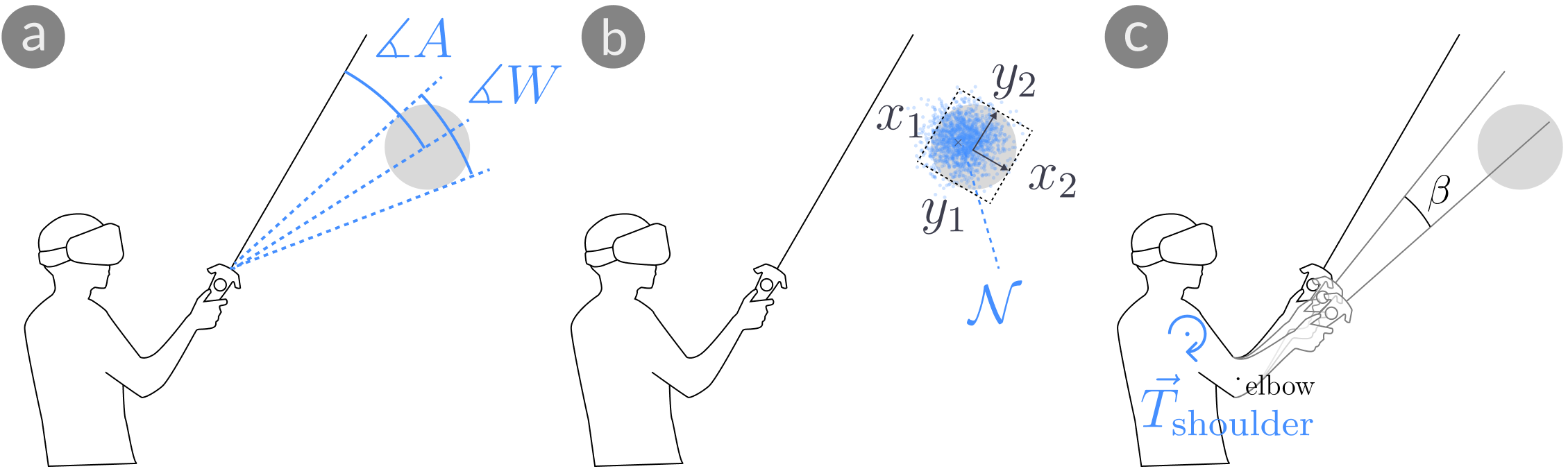}
    \caption{Scoring parameters used for objectives (a) Speed, (b) Accuracy, and (c) Comfort. The grey sphere is the target.}
    \Description{This image is a three-part illustration (labeled a, b, and c) showing scoring parameters used for evaluating Speed, Accuracy, and Comfort. In all figures, a person wearing a VR headset is aiming a controller toward a gray target sphere. 
    Figure (a) shows two angular measurements, ∠A and ∠W. ∠A represents the angular distance from the pointing direction to the center of the target, while ∠W represents the angular width of the target. 
    Figure (b) depicts the endpoint distribution of the selection task, which follows a normal distribution, labeled N. The bounding box around the target has coordinates (x1,y1,x2,y2), representing its boundaries. 
    Figure (c) shows the torque applied at the shoulder, labeled T_{shoulder}, and a sequence of hand postures pivoting at the elbow, rotating incrementally by an angle β toward the target center.
    }
    \label{fig:scoring}
\end{figure}

Here, $\vec{r}_\text{arm}$ is the distance from the shoulder joint to the center of mass of the arm, and $m$ is the mass of the arm. Since users need to move their arms to reach a target, we calculate the shoulder torque based on predicted user poses during pointing.
For simplicity, we assume that users will rotate their forearm toward the aiming center of the target (as defined in \autoref{sec:objective-speed}) at a constant rotational speed, keeping the elbow fixed and aligning the ray with the forearm. While simplified, this approach captures overall fatigue trends and is robust to inaccuracies due to \system{}'s continuous score recalculation as users adjust posture.
Furthermore, we know that longer interaction time requires more energy and can lead to fatigue when endurance limits are reached~\cite{hincapie-ramos2014Consumed}. To quantify this effect of time in energy, we sum shoulder torques from user poses along the pointing trajectory, sampled at $\beta$-degree increments from the initial posture (defined in \autoref{sec:context-information}) to the final target orientation (\autoref{fig:scoring}c). This total exertion is then negated to yield the comfort score:

\begin{equation}
    S_C = -\sum_{\text{pos}_i}||\vec{T}_{\text{shoulder}, \text{pos}_i}||.
    \label{eq:comfort}
\end{equation}

\subsubsection{Familiarity}
Although an advanced technique may be more efficient according to the defined objectives, users may prefer simpler techniques when sufficient to reduce effort and cognitive load. To capture this, each technique is assigned a \emph{``familiarity''} score:

\begin{equation}
    S_F = S_{F\text{, }tech} \text{  if Technique = } tech.
    \label{eq:familiarity}
\end{equation}

This score reflects the added complexity associated with advanced selection techniques and how they may limit interaction in qualitative ways, which is a trade-off between performance and usability.
For example, a technique designed for occluded selection may rely on more interaction steps than normal pointing and include mechanisms that alter the typical pointing user experience by, for example, only supporting discrete selection and not continuous targeting~\cite{baloup2019RayCursor, molina2023Twostep}.
Therefore, simpler techniques that align more closely with traditional pointing are expected to have higher familiarity scores than advanced techniques that require more interaction steps~\cite{scarr2011Dips}. In our implementation, we give each candidate technique a constant value $S_{F\text{, }tech}$ based on pilot testing in simple selection tasks. In the future, we envision the potential for individual adaptation based on user exposure and performance with each technique~\cite{octavia2011Adaptation}. 

\subsubsection{Normalization and Aggregation}
To ensure that objective scores are comparable and measured on a consistent scale, we normalize these scores using Min-Max normalization, where the minimum ($s_{min}$) and maximum ($s_{max}$) represent the theoretical limits of each objective model. We consider extreme cases in our implementation for limits that do not have a theoretical bound. For example, we consider the smallest target size based on the display limitations and the largest target amplitude as the angle of the interaction region cone ($r_c$), and the largest possible motion at the most strenuous user position. For special cases such as when an object's activation region is zero due to occlusion, we assign the minimum value as the target is unselectable. To reduce noise caused by environmental or user factors, we apply an exponential smoothing factor to all objectives calculated for each target and technique, defined by a smoothing factor $\alpha$. Therefore, the score of each objective in the time frame $t$ is defined as

\begin{equation}
    S_t = \alpha \times \frac{\sum{s_{\text{obj}_i}} / N - s_{min}}{s_{max} - s_{min}} + (1-\alpha) \times S_{t-1}.
    \label{eq:smoothing}
\end{equation}

To aggregate the effects of all objects within the interaction space and into a single representative value, we calculate the average of each objective score across all objects within the interaction space. This implies that each object is treated equally important for optimization. In future versions, weighted averages together with target prediction approaches~\cite{henrikson2020HeadCoupled, Chung2024Prediction} could be deployed to give objects that are more likely to interact with a higher priority.

\subsection{Technique Switching} \label{sec:adaptation-and-switching}
To consider the comprehensive results of the objectives for deciding the most optimal selection technique, we calculate a weighted sum of the scores and select the one with the maximum overall score:
\begin{equation}
    \text{Optimal} = \operatorname*{argmax}_{tech} \left( k_S \times S_S + k_A \times S_A + k_C \times S_C + k_F \times S_F \right).
    \label{eq:optimization}
\end{equation} 
Designers can give the objectives different weightings ($k$) depending on user tasks or contexts. For example, in a password input task, designers might want to prioritize accuracy and therefore give a higher weight to the accuracy objective. 

Finally, to activate a switch in the interaction technique, we ensure that the optimal technique is optimal for $n$ frames within a $w$-frame window to ignore brief and sudden technique switches. Within these $n$ frames, the difference between the most optimal technique and the current technique must be greater than $t_o$. Although this introduces a minor delay, we posit that users will be more susceptible to technique switches if the switches are only performed when needed. $w$, $n$ and $t_o$ can be adjusted to tune the responsiveness and sensitivity of switching. We applied haptic and audio feedback to help users notice technique switches. The controller and ray also change to a unique color for visual feedback. These design choices aim to mitigate distraction and preserve user awareness in an automatic switching system.

\subsection{Selection Techniques} \label{sec:technique}
A significant number of selection techniques have been proposed for XR~\cite{argelaguet2013Survey} and the selection techniques included in the optimization will have a significant impact on the \system{} user experience. We chose selection techniques based on a set of criteria and assumptions: (1) the techniques should be pointing-based and the working mechanism should alter the target width and amplitude to be compatible with our implemented objectives; (2) the techniques should cover a range of selection scenarios to benefit from adaptation and switching; (3) we should limit the number of techniques to only one per selection scenario to limit the required user training and confusion due to switching between a large set of techniques; and (4) the techniques should not require extra sensing, hardware, or modalities beyond a typical VR controller.
Our version of \system{} is implemented with three controller-based pointing techniques: \textit{RayCasting}, \textit{StickyRay}, and \textit{RayCursor}. These were chosen to cover a wide range of pointing scenarios, from regular pointing to small targets, dense environments, and occlusion. 
Although our current implementation focuses on these three techniques, \system{} can be extended to incorporate additional techniques. To integrate these techniques into \system{}, the selection technique must define $W$, $A$ and $(x_1, x_2, y_1, y_2)$ for every target in the interaction space. Each technique must also define $S_{F_{i}}$ for the familiarity objective. In the following section, we introduce each technique included and explain how these variables are derived to support objective computation.

\begin{figure}
    \centering
    \includegraphics[width=0.8\linewidth]{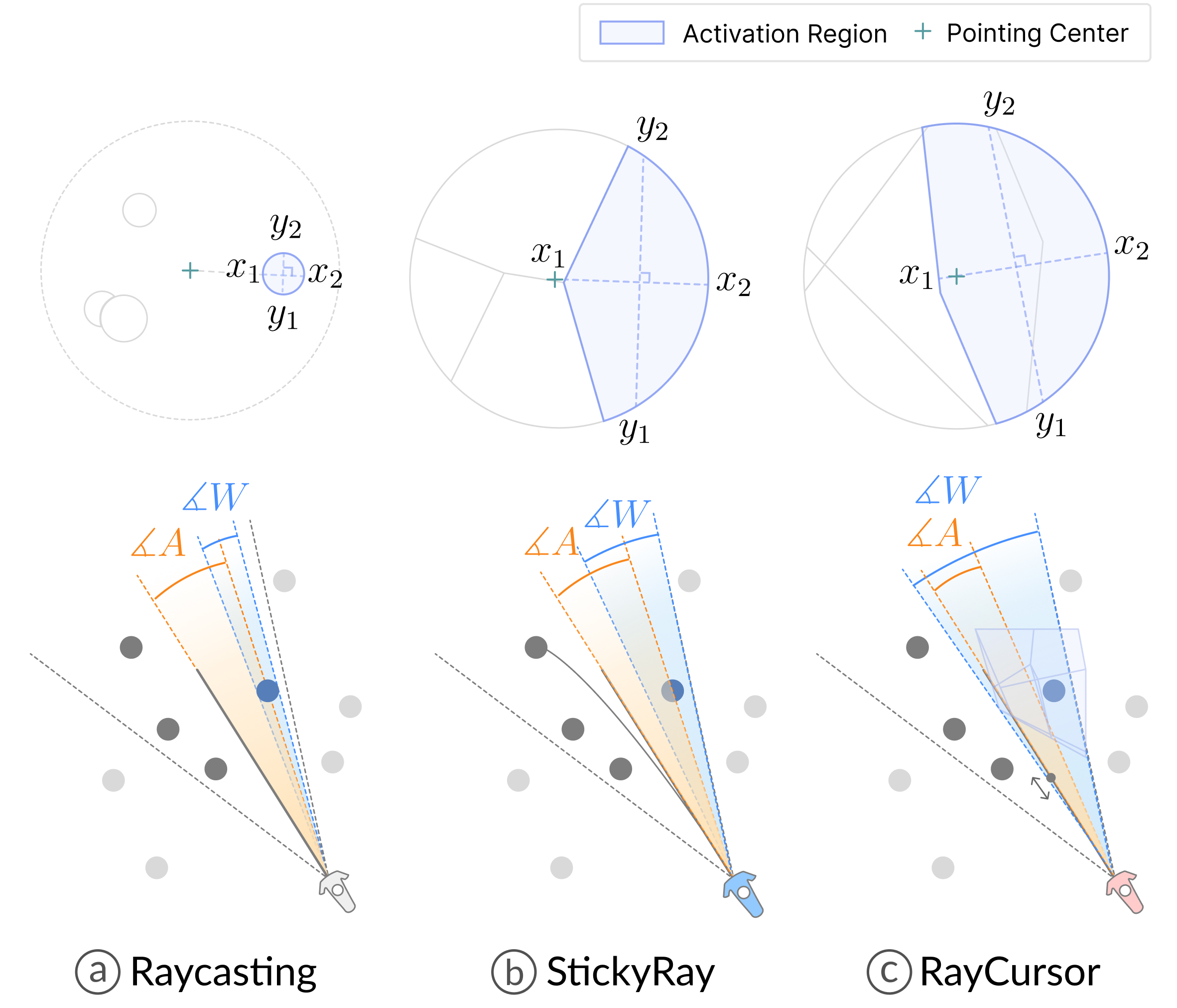}
    \caption{Context score implemented for 3 techniques (a) RayCasting, (b) StickyRay, and (c) RayCursor. The lower figure visualizes the effective size ($W$) and amplitude ($A$) of the blue target. The upper figure shows the projection of the activation region of the target.}
    \Description{This image consists of two rows illustrating the Context score implemented for three techniques: (a) RayCasting, (b) StickyRay, and (c) RayCursor. The lower row visualizes the raycasting with a cone of interaction space, while the upper row shows the projection of the activation region from the controller. Gray peripheral targets and a blue target are used for illustration. The parameters of the blue target are depicted: the lower row shows the effective size (∠W) and amplitude (∠A) while the upper row shows the coordinates (x1, x2, y1, y2) marking the boundaries of its interaction, along with the centroid of the interaction region. 
    Figure (a) RayCasting: In the lower row, the effective width ∠W is shown as angular measurements between the closest and furthest edges of the blue target, while amplitude∠A represents the angular distance between the controller’s pointing direction and the closest edge of the target. In the upper row, the target’s activation region matches the object's circular area with coordinates (x1, x2, y1, y2) defining the target’s boundaries. 
    Figure (b) StickyRay: In the lower row, the controller points at the target using the StickyRay technique, where the ray bends toward the nearest target. Angles ∠W and ∠A are shown similarly to RayCasting, but the interaction area is larger, corresponding to the 2D Voronoi cell of the blue target. In the upper row, the activation region is represented as a 2D Voronoi diagram, with the blue target’s activation region and coordinates (x1, x2, y1, y2) showing a larger interaction region than in RayCasting.  
    Figure (c) RayCursor: In the lower row, the RayCursor technique introduces a cursor along the ray, with a 3D Voronoi cell surrounding the blue target, displayed as a crystal-like structure. The angles ∠W and ∠A for the 3D Voronoi cell, which represent the interaction space. In the upper row, the activation region is projected from the 3D Voronoi diagram as the interaction space, and the coordinates (x1, x2, y1, y2) again define the target’s interaction boundaries.
    }
    \label{fig:region}
\end{figure}

\subsubsection{RayCasting}
RayCasting represents the most basic pointing technique in which the user points with a ray originating from the controller and the target hit by the ray is highlighted for selection. Due to its simplicity and popularity in 3D interaction, we treat it as a base case for selection. We define $A$ as the angular distance from the current pointing direction to the target center in 2D space (\autoref{sec:context-information}), and $W$ as the angular width of the object along the pointing direction; $x_1$ and $x_2$ denote the entry and exit points of the 2D object along the pointing direction, while $y_1$ and $y_2$ correspond to the intersections of the object's outline with a line perpendicular to the pointing direction and passing through the target center (\autoref{fig:region}a). For completely occluded targets (i.e., those with no visible activation region in the 2D projection due to occlusion) we assign the minimum score.
However, users may change their point of view to reveal such targets, and the system reflects these changes as it continuously adjusts in real time during selection.
Despite RayCasting being the most widely used technique, users' inherent hand tremors can result in instability in pointing accuracy, particularly when selecting small objects. Additionally, when the number of objects in the environment increases, selection becomes more difficult due to the close proximity of objects and occlusion.



\subsubsection{StickyRay}
We included StickyRay as a second technique for small target selection. StickyRay is based on the Bubble Cursor metaphor~\cite{grossman2005Bubble}, where the object nearest to the pointing direction is highlighted for selection. This mechanism expands the effective width of each target to a region that together builds a Voronoi diagram, thus making easy selection of small targets. To show the current closest object, a secondary ray bends toward it~\cite{steinicke2006Object}. We used the angular distances from the pointing direction to the targets to decide the current closest target, as it has been shown to be the best performing version in 3D settings~\cite{lu2020Investigating}. As such, the object activation region is the space in which the ray forms the smallest angular distance to the object compared to all other objects (\autoref{fig:region}b). This is equivalent to computing a Voronoi region in the projected 2D space, where distances are measured by angular distance to the ray. We use Qhull~\cite{barber1996Quickhull} to find the 2D Voronoi region and then clip it by the interaction space. We first find the two intersection points $p_1$ and $p_2$ of the line of movement and its projected 2D Voronoi region. Then we get $A$ as the angular distance from the origin to the centroid point of the Voronoi region, and $W$ as the angular distance between $p_1$ and $p_2$. $x_1$, $x_2$, $y_1$, and $y_2$ are defined as with RayCasting but instead using the Voronoi region as target borders.

While it is easier to select small targets compared to RayCasting, StickyRay can be unintuitive, as it encourages pointing outside the visual boundaries of the target. Furthermore, a consequence of the Bubble Cursor mechanism is that a target will always be highlighted for selection, which may not be preferable depending on the context of use. Finally, although StickyRay is proficient in selecting small targets and selection in sparse environments, its benefit diminishes in crowded environments and occluded targets. 

\subsubsection{RayCursor}
To handle dense environments and target occlusion, we included RayCursor, where the user controls a cursor on the ray by swiping on the controller touchpad to select targets at different depths~\cite{baloup2019RayCursor}. Like StickyRay, RayCursor has a proximity selection mechanism that will pre-select the object nearest to the cursor to improve performance in selecting small or distant targets. To minimize the need for swiping, the technique has a snapping mechanism that moves the cursor immediately to the depth of the first pointed object's surface. We chose the semi-auto version of RayCursor with the VitLerp transfer function for cursor movement, as it was shown to be the highest performing version~\cite{baloup2019RayCursor}. The semi-auto version disables the snapping mechanism when users manually control the cursor through swiping, and reactivates after the trackpad has been released for more than one second. Since RayCursor allows selection either by ray movement alone or in combination with trackpad swiping, its movement is difficult to define. Therefore, we only consider the controller movement for modeling to simplify the calculations. To calculate $A$ and $W$, we first compute a 3D Voronoi region based on the provided 3D space using QHull~\cite{barber1996Quickhull}. We then project the 3D Voronoi regions to the control space and calculate $A$ and $W$ as in other techniques while ignoring occlusion. $x_1$, $x_2$, $y_1$, and $y_2$ are defined as with RayCasting and StickyRay but instead using 3D Voronoi projected to control space~ (\autoref{fig:region}c).

The RayCursor provides easier selection in dense and occluded environments as it leverages additional depth information for selection. However, the additional interaction steps necessary increase its complexity compared to RayCasting and StickyRay.




\subsection{Implementation} \label{sec:implementation}
We implemented \system{} in Unity. We used the HTC Vive Pro Eye which has a $110^\circ$ FOV and a $2880 \times 1600$ resolution and the Vive controller for pointing input. The controller trackpad was used to control the cursor for RayCursor and the trigger was used to select targets. We used the built-in Unity Inverse Kinematics library to generate user postures. We relied on previous studies to define values for objective parameters that require empirical values. For the accuracy objective, we relied on previous studies by \citet{yu2019Modeling} to establish values for the endpoint distribution model: $\mu = -0.1441 \times W + 0.2649$, $\sigma_x = 0.0066 \times A + 0.1025 \times W + 0.2663$, and $\sigma_y = 0.0085 \times A + 0.0679 \times W + 0.1437$. For the comfort objective, we used the equation specified by ~\citet{hincapie-ramos2014Consumed} to calculate the center of mass for r and $m$. As for the input data for body parameters we used the average human data for simplicity, specified by \citet{freivalds2011Biomechanics}:
33cm long upper arm weighing 2.1 kg with the center of mass located at 13.2cm; 26.9cm long forearm weighing 1.2 kg with the center of mass located at 11.7cm; 19.1cm long hand weighing 0.4kg with the center of mass located at 7cm.

\section{Application} \label{sec:application}
\begin{figure*}[t]
    \centering
    \includegraphics[width=.9\linewidth]{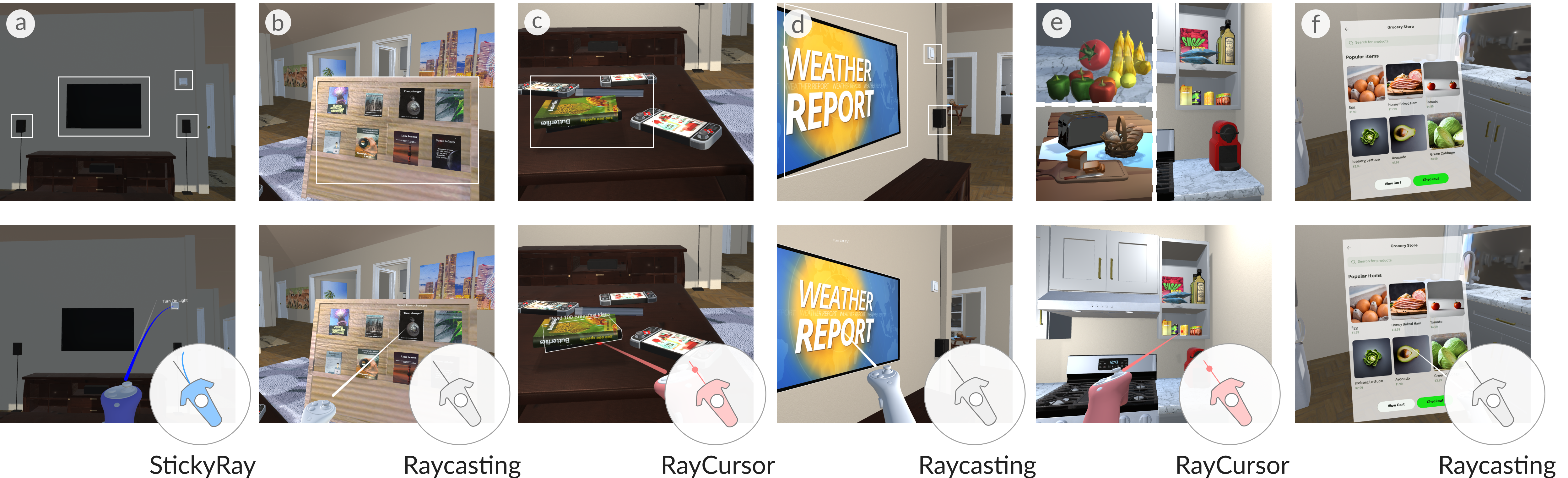}
    \caption{\system{} chooses different techniques for different scenarios in the application, such as interacting with (a) IOTs on the far wall, (b) books layout on the nearby bookshelf, (c) books stacked on the coffee table, (d) IOTs on the side, (e) ingredients in the cluttered kitchen, and (f) UI panel in front of the users.}
    \Description{This image is divided into two rows, demonstrating interaction techniques used in a virtual environment. In the upper row, the environment is shown with possible interaction targets circled, while in the lower row, the interactions and the specific techniques used for selection are shown. In the upper row: figure (a) shows a living room wall with speakers, a TV, and a light switch; figure (b) displays books on a shelf in a room; figure (c) shows books stacked on a table along with game controllers; figure (d) shows a TV and a light switch on a wall; figure (e) shows a kitchen counter cluttered with various food and kitchenware such as fruits, toast, and cans; figure (f) shows a UI panel of an online grocery app with food items and selection buttons with pictures. In the lower row: figure (a) shows the StickyRay technique being used to point at the light switch on the far wall; figure (b) shows the RayCasting technique pointing at a book on the shelf; figure (c) shows the RayCursor technique being used to point at a book in a stack on the table; figure (d) shows the RayCasting technique pointing at the TV; figure (e) shows the RayCursor technique being used to select a food item from the upper cabinet in the kitchen; figure (f) shows the RayCasting technique pointing at a button in the grocery app UI panel.}
    \label{fig:demo}
\end{figure*}

To show \system{}'s performance in dynamic settings, we developed an indoor VR environment where users interact with IOTs, books, food, and UI elements that have different sizes, arrangements, and density. \system{} is designed to responsively and smoothly switch the selection technique as the user's focus shifts.

In this app, users can point to an interactable target, which displays a brief command description (e.g., `turn on the light'). By pressing the trigger button on the controller, the command is executed. The application and its interactions are designed to represent common selection scenarios found in VR environments. \system{} was developed as specified in \autoref{sec:system}. We applied the following objective weightings: $k_S = 0.5$ for speed, $k_A = 0.2$ for accuracy, $k_C = 0.15$ for comfort, and $k_F = 0.15$ for familiarity. We applied the following normalized familiarity scores: $S_{F,\textit{ RayCasting}} = 0.57$, $S_{F,\textit{ StickyRay}} = 0.33$, $S_{F,\textit{ RayCursor}} = 0.1$, and the following parameters: interaction space cone with radius $r_c=20^\circ$, smoothing factor $\alpha=0.8$, the threshold of $n=15$ number of frames with improvement above $t_{o} = 0$ within $w=20$ windows, and the rotational increments of $\beta=1^\circ$ to derive interaction postures. These parameters are chosen based on informal pilot testing to ensure the objectives are balanced, the switching is not too sensitive, and no ambiguity on the decision of the most optimal technique.

To exemplify the benefits of \system{}, we detail a walkthrough of the application (see Video Figure). In the living room of the virtual house, the user first points to the light switch on the wall, attempting to turn on the light. Since it is too far away and small, they struggle to select at the beginning. \system{} continuously senses the environment and user's state and immediately switches RayCasting to StickyRay. StickyRay snaps the ray to the light switch and makes the selection easier (\autoref{fig:demo}a). Afterward, they want to pick a book to read. There are books laid out on the shelves and stacked on the table. \system{} smoothly switches to RayCasting when the books are large enough for easy selection (\autoref{fig:demo}b), and switches to RayCursor when the book is occluded (\autoref{fig:demo}c). The user controls the depth of the cursor on the trackpad to pick the book hidden behind. With the responsive assistance of \system{}, they can precisely select the book to read. They select a sandwich-making guide and decide to go to the kitchen to check the required ingredients. On their way to the kitchen, they turn off the light and TV to save electricity. \system{} chooses the simplest RayCasting because the IOTs are near and big, thus easy to select~(\autoref{fig:demo}d).

The kitchen is cluttered with food and kitchenware. \system{} chooses RayCursor to handle the dense environment (\autoref{fig:demo}e). This allows them to easily pick the tomato in the stack of fruits, the piece of toast on the cutting board with a blanket of bread around, and the bottle of olive oil arranged on the top cabinet. After they check all the food ingredients, they find that ham and cabbage are out of stock. Therefore, they open an online grocery shopping app to order them.  Since the UI buttons on the pop-up panel are designed for easy interaction, \system{} picks RayCasting for easy interaction (\autoref{fig:demo}f). They select the ingredients icon and the checkout button to order and wait for the ingredients to be delivered. In sum, the application showcases the following advantages:
\begin{itemize}
    \item \system{} responsively switches the technique when the content of the user's interest changes, assisting users to interact with a non-homogeneous environment.
    \item \system{} comprehensively considers the user’s performance in time, accuracy, comfort, and familiarity. When the task is easy enough to be used with a basic technique, \system{} will stick to the basic one. When the task becomes harder to complete with that technique, it will automatically switch to a more advanced and suitable one.
    \item \system{} provides a smooth, consistent, and non-distracting transition by proactively switching the selection tool before users point toward new targets and ensuring that the tool remains consistent when engaging with nearby objects. 
\end{itemize}
\newcommand{\anova}[6]{($F_{#1,#2}$$=$$#3$, $p$$#4$$#5$, $\eta_{p}^{2}$$=$$#6$)}
\newcommand{\ph}[2]{$p$$#1$$#2$}

\section{Evaluation}
We conducted a VR user study to explore and evaluate the potential quantitative benefits of \system{} across different scenarios, compared to using individual techniques. We hypothesized that relying on a single selection technique would introduce performance trade-offs whereas Adaptique always achieve optimal performance.

The study was done through a controlled task where participants were instructed to select one target among many distraction targets as quickly and accurately as possible in different environments. We collected performance metrics, including selection time, error rate, translational movement, and rotational movement. These correspond to our core performance objectives: selection time reflects speed, error rate reflects accuracy, and movement reflects ergonomic cost. To test our hypothesis, we evaluated \system{} using StickyRay and RayCursor as selection techniques. 
Raycasting was not included in this performance-based study because, theoretically, its effective size is a lower bound and its effective amplitude is an upper bound, making it inherently outperformed by other techniques on performance metrics~\cite{baloup2019RayCursor, lu2020Investigating}, and thus unlikely to offer additional insight into the performance trade-offs we aimed to evaluate.



\subsection{Task}
\begin{figure}
    \centering
    \includegraphics[width=.8\linewidth]{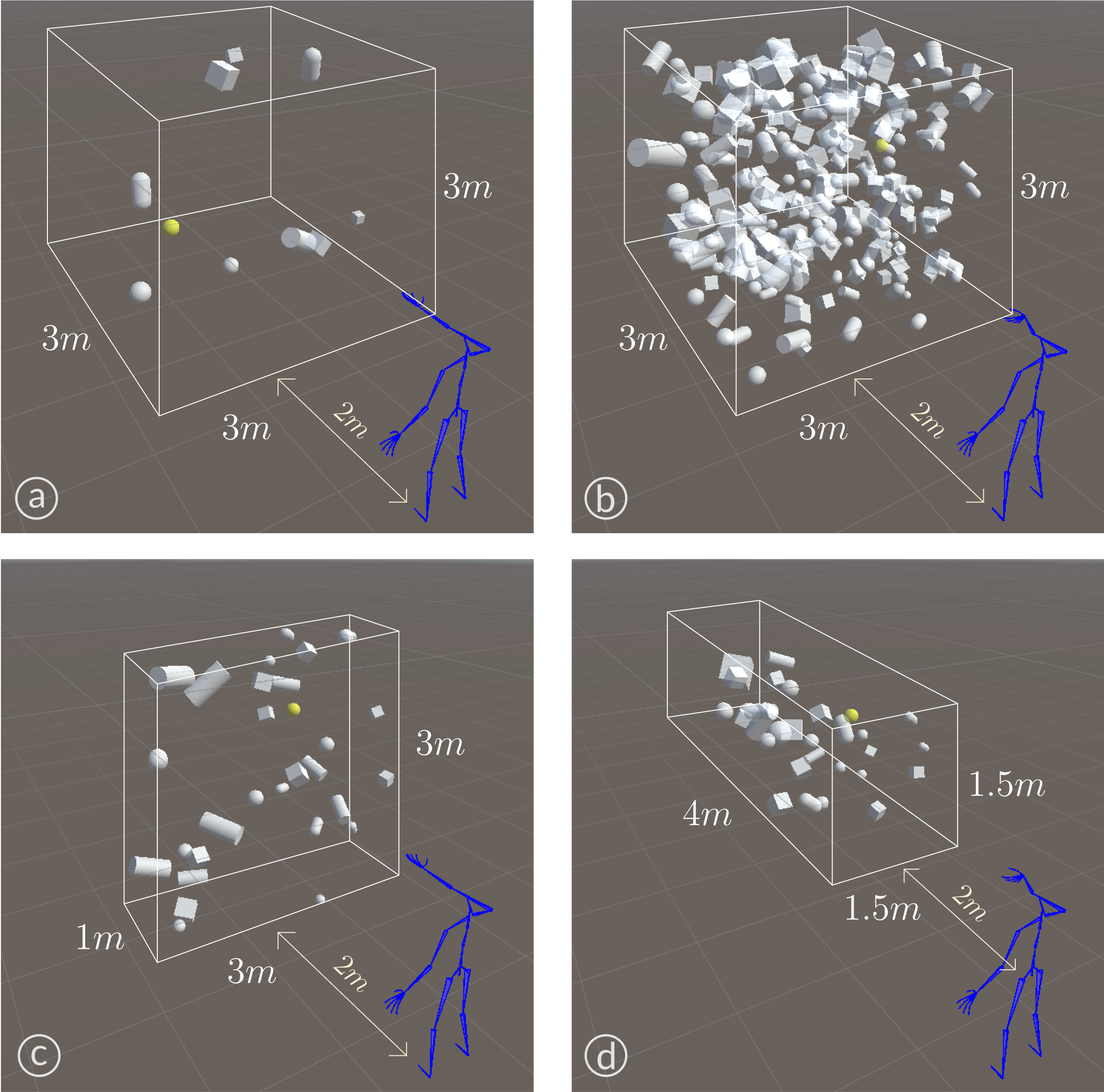}
    \caption{Environments used in the study, including (a) Sparse (b) Dense (c) Flat, and (d) Deep.}
    \Description{This image consists of four 3D visualizations labeled (a), (b), (c), and (d), showing different target environments for a selection task in XR. In each figure, the participant is represented by a skeletal figure standing 2 meters away from a cubic or rectangular target region. The region contains one yellow sphere (the target object) and various white distractor objects, including cubes, spheres, cylinders, and capsules. The size of the cubic/rectangular region and the number of distractor objects vary across the environments. 
    Figure (a) Sparse Environment: A cubic region measuring 3m×3m×3m is shown with 10 objects inside, including 1 yellow target sphere and 9 white distractors. The objects are spaced far apart, representing a simple, sparse selection scenario. 
    Figure (b) Dense Environment: A cubic region of the same size (3m×3m×3m) contains 240 distractor objects densely packed around the yellow target sphere. The distractors partly occlude the target, creating a challenging, dense selection environment. 
    Figure (c) Flat Environment: A flattened cubic region measuring 3m×3m×1m is displayed with 30 objects. The objects are spread out at the same depth, with the yellow target sphere positioned among 29 white distractors. This environment presents a horizontally spread selection task. 
    Figure (d) Deep Environment: A deep rectangular region measuring 1.5m×1.5m×4m contains 30 objects. The objects are arranged to extend more into the depth of the space, with the yellow target sphere surrounded by 29 white distractors. This environment emphasizes selection in a more elongated, depth-oriented space.
    }
    \label{fig:study-env}
\end{figure}

Participants were tasked to select a target object amongst many distraction targets. 
We varied the size of the selectable target, the number of targets, and the density of the environment. Target sizes were specified in visual angles, with two conditions: large (2.5$^\circ$) and small (0.5$^\circ$), and four different environments that varied in distractors amount and density to cover both simple and extreme cases that users might encounter in XR. The selection target was a sphere and randomly placed within the target region but had to be at least 0.4 meters away from the boundary so that the target was not at the edge within the environment and 0.2 meters away from the center to ensure movement before selection. The distractors were primitive shapes (cubes, spheres, cylinders, and capsules), in pseudo-random positions and sizes (2-4$^\circ$), and random rotations. They were semitransparent to minimize the effect of visual search in dense and occluded settings and were not intersecting with each other.
The environments were also balanced, so that two environment types exploited the advantages of StickyRay, and similarly two environments exploited the advantages of RayCursor. Specifically, the environments were as follows: 

\begin{description}
\item[Sparse:] In \textsc{Sparse} environment (\autoref{fig:study-env}a), the target object and distractors were spawned within a $3m\times3m\times3m$ cubic region 2 meters away from the participant. There were a total of 10 objects: 1 target object and 9 distractors. The environment represented a simple case of selection.
\item[Dense:] The \textsc{Dense} environment (\autoref{fig:study-env}b) also consisted of a $3m\times3m\times3m$ cubic region 2 meters away but contained 240 objects, making the selection target densely surrounded by other objects. The target was likely to be partly occluded by distractors from the view of the participant.
\item[Flat:] In \textsc{Flat} environment (\autoref{fig:study-env}c), the spawning region was a $3m\times3m\times1m$ cubic region 2 meters away, resulting in a spread-out placement at a similar depth. We used a total of 30 objects.
\item[Deep:] In \textsc{Deep} environment (\autoref{fig:study-env}d), the spawning region was a $1.5m\times1.5m\times4m$ cubic region 2 meters away. A total of 30 objects were spawned. Though the density of the environment (30 objects in $90m^3$ volume) was the same as \textsc{Flat} environment, the arrangement of objects extended more in the depth direction.
\end{description}

We pre-generated eight trials of each unique combination of target size and environment to use for all techniques. In sum, the study used the following independent variables and levels:
\begin{itemize}
    \item \textsc{Technique: StickyRay, RayCursor, Adaptique}
    \item \textsc{Target size: Small ($\SI{0.5}{\degree}$), Large ($\SI{2.5}{\degree}$)}
    \item \textsc{Target environment: Sparse, Dense, Flat, Deep}
\end{itemize}


\subsection{Procedure}
Upon arrival, participants completed a consent form and a demographic questionnaire to gather information about participant age, gender, and VR/AR experience before being informed about the study. They were then positioned in a standing posture, equipped with the HMD and controller, and performed a practice session to familiarize themselves with each selection technique before starting the study session. During the study session, the participants performed all selections with a single technique before moving on to the next technique. For each technique, a total of 64 trials (2 \textsc{Target size} $\times$ 4 \textsc{Target environment} $\times$ 8 repetitions) were presented in random order. The order of the techniques was counterbalanced with a balanced Latin square. For each trial, participants needed to select a central ``ready'' panel before starting the trial. This served as a rest period and ensured that users started the next trial from a central position. Participants were then tasked to select the target object that was highlighted in yellow as quickly and accurately as possible. The participants were unable to move on to the next trial until the correct object had been selected or a 15-second timeout had elapsed. After finishing all the trials with a technique, participants completed a questionnaire consisting of Likert-scale usability questions to capture their experiences. Study usability questions can be found in the supplemental material. Participants then rested before moving on to the next technique. 
The study was concluded with a questionnaire for user preferences and feedback. In total, we collected 18 participants $\times$ 3 \textsc{Technique} $\times$ 2 \textsc{Target size} $\times$ 4 \textsc{Target environment} $\times$ 8 repetitions = 3456 selections.

\subsection{Apparatus and Participants}
The interaction techniques and \system{} were implemented as described in \autoref{sec:system}. The participants performed the tasks with the controller in their dominant hand. Selection was done with the trigger button, and depth cursor control with the trackpad. Since our study focuses on performance, we applied the following objective weightings: $k_S = 0.5$, $k_A = 0.2$, $k_C = 0.2$, and $k_F = 0.1$. We applied the following normalized familiarity scores: $F_{\text{StickyRay}} = 0.7$ and $F_{\text{RayCursor}} = 0.3$ based on pilot testing. The rest of the parameters are the same as those in ~\autoref{sec:application}. 
We recruited 18 participants on campus for the study (12 male, 6 female, 19-32 years old). One used VR/AR weekly, thirteen used VR/AR occasionally, and four had never experienced VR/AR before.

\subsection{Results}
Unless otherwise stated, the analysis was performed with a 3-way repeated measures ANOVA ($\alpha$=$.05$) with \textsc{Technique}, \textsc{Size}, and \textsc{Environment} as independent variables. Before analysis, we removed outlier trials. Trials were discarded if their selection times, translational movement, or rotational movement were beyond 3 standard deviations from their respective grand mean. In total, 164 out of 3456 trials were discarded (4.7\%). We tested normality with the Kolmogorov-Smirnov test and QQ-plots. If extreme outliers were identified within the aggregated analysis data, defined as values beyond $Q_R \pm 3 \times \text{IQR}$, a winsorization process was applied. When the assumption of sphericity was violated, as tested with Mauchly’s test, Greenhouse-Geisser corrected values were used in the analysis. Bonferroni-corrected post hoc tests were used when applicable. The effect sizes were reported as partial eta squared ($\eta_{p}^{2}$). Questionnaires were analyzed using Friedman tests, and Bonferroni-corrected Wilcoxon signed-rank tests for post hoc analysis. 
\begin{figure*}[t]
    \centering
    \includegraphics[width=.9\linewidth]{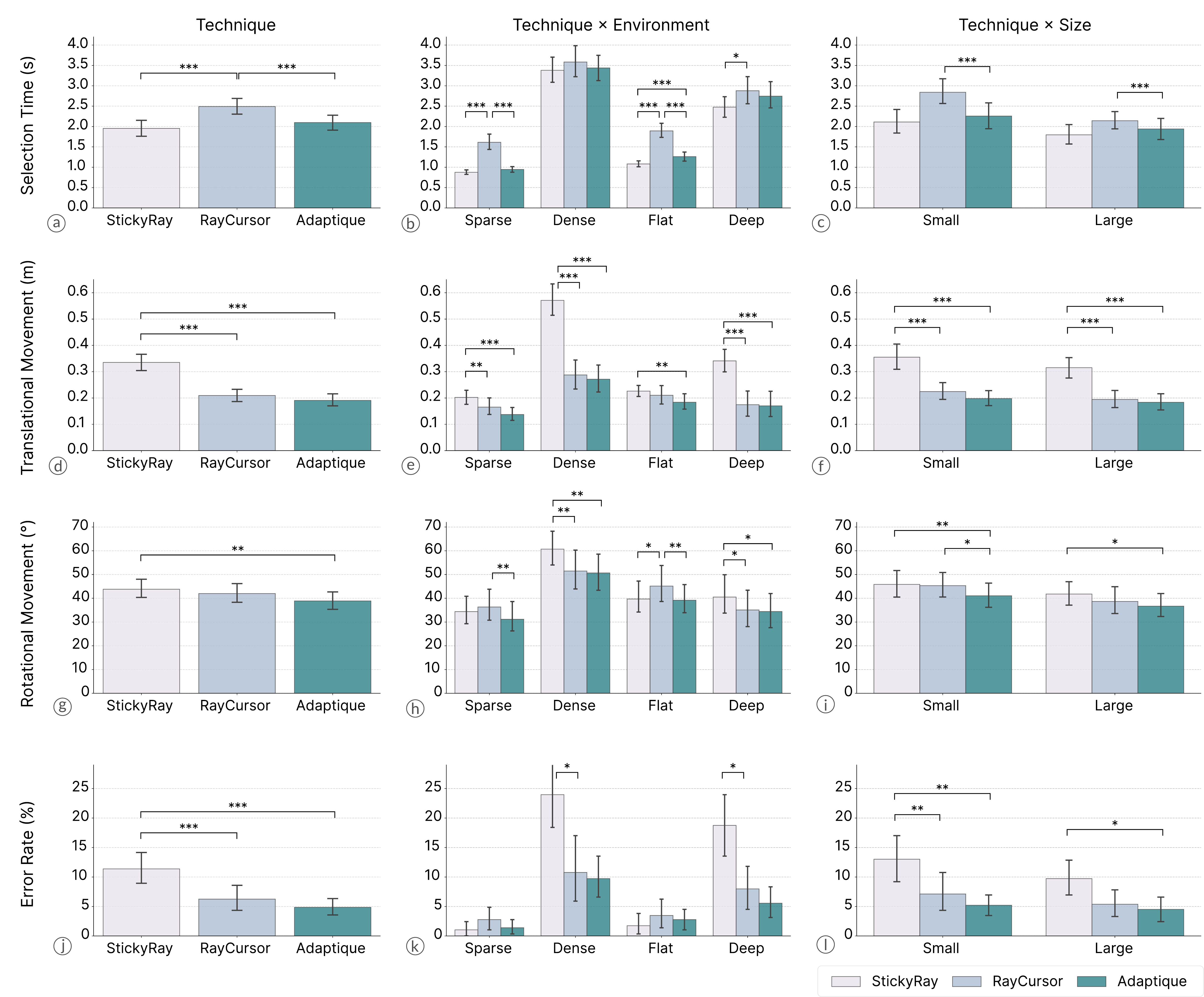}
    \caption{Mean selection time, translational movement, rotational movement, and error rate. Error bars represent the mean 95\% confidence intervals. The symbol $*$ indicates \ph{<}{.05}, $*$$*$ indicates \ph{\le}{.01}, and $*$$*$$*$ indicates \ph{\le}{.001}.}
    \Description{The figure contains multiple bar plots with error bars representing the mean and variance for selection time, translational movement, rotational movement, and error rate across three interaction techniques (StickyRay, RayCursor, Adaptique), compared under various conditions (environment and size); significant differences are indicated by * (p ≤ 0.05), ** (p ≤ 0.01), and *** (p ≤ 0.001); 
    (a) Selection Time by Technique: StickyRay: mean 1.95 s, variance 1.44, RayCursor: mean 2.49 s, variance 1.39, Adaptique: mean 2.10 s, variance 1.57, *** between StickyRay and RayCursor, and RayCursor and Adaptique; 
    (b) Selection Time by Technique and Environment: StickyRay Sparse: mean 0.88 s, variance 0.03, RayCursor Sparse: mean 1.61 s, variance 0.35, Adaptique Sparse: mean 0.95 s, variance 0.04, StickyRay Dense: mean 3.38 s, variance 0.85, RayCursor Dense: mean 3.58 s, variance 1.40, Adaptique Dense: mean 3.44 s, variance 0.91, StickyRay Flat: mean 1.08 s, variance 0.05, RayCursor Flat: mean 1.89 s, variance 0.29, Adaptique Flat: mean 1.26 s, variance 0.12, StickyRay Deep: mean 2.48 s, variance 0.62, RayCursor Deep: mean 2.88 s, variance 1.07, Adaptique Deep: mean 2.75 s, variance 0.96, significant differences: Sparse (), Flat ( for all pairs), Deep (* between StickyRay and RayCursor); 
    (c) Selection Time by Technique and Size: StickyRay Small: mean 2.11 s, variance 1.75, RayCursor Small: mean 2.84 s, variance 1.63, Adaptique Small: mean 2.26 s, variance 1.89, StickyRay Large: mean 1.80 s, variance 1.11, RayCursor Large: mean 2.14 s, variance 0.92, Adaptique Large: mean 1.94 s, variance 1.21, *** for small and large sizes between StickyRay and Adaptique; 
    (d) Translational Movement by Technique: StickyRay: mean 0.34 m, variance 0.04, RayCursor: mean 0.21 m, variance 0.02, Adaptique: mean 0.19 m, variance 0.02, *** between StickyRay and RayCursor, StickyRay and Adaptique; 
    (e) Translational Movement by Technique and Environment: StickyRay Sparse: mean 0.20 m, variance 0.01, RayCursor Sparse: mean 0.17 m, variance 0.01, Adaptique Sparse: mean 0.14 m, variance 0.01, StickyRay Dense: mean 0.57 m, variance 0.03, RayCursor Dense: mean 0.29 m, variance 0.03, Adaptique Dense: mean 0.27 m, variance 0.03, StickyRay Flat: mean 0.23 m, variance 0.01, RayCursor Flat: mean 0.21 m, variance 0.01, Adaptique Flat: mean 0.18 m, variance 0.01, StickyRay Deep: mean 0.34 m, variance 0.02, RayCursor Deep: mean 0.17 m, variance 0.02, Adaptique Deep: mean 0.17 m, variance 0.02, significant differences: Sparse (*** between StickyRay and Adaptique, ** between StickyRay and RayCursor), Dense (*** StickyRay and Adaptique), Flat (** StickyRay and Adaptique), Deep (*** StickyRay and RayCursor, StickyRay and Adaptique); 
    (f) Translational Movement by Technique and Size: StickyRay Small: mean 0.36 m, variance 0.04, RayCursor Small: mean 0.22 m, variance 0.02, Adaptique Small: mean 0.20 m, variance 0.02, StickyRay Large: mean 0.32 m, variance 0.03, RayCursor Large: mean 0.19 m, variance 0.02, Adaptique Large: mean 0.18 m, variance 0.02, *** for small and large sizes between StickyRay and RayCursor; 
    (g) Rotational Movement by Technique: StickyRay: mean 43.82°, variance 552, RayCursor: mean 42.00°, variance 576, Adaptique: mean 38.87°, variance 495, ** between StickyRay and Adaptique; 
    (h) Rotational Movement by Technique and Environment: StickyRay Sparse: mean 34.38°, variance 308, RayCursor Sparse: mean 36.32°, variance 387, Adaptique Sparse: mean 31.17°, variance 319, StickyRay Dense: mean 60.71°, variance 496, RayCursor Dense: mean 51.47°, variance 634, Adaptique Dense: mean 50.67°, variance 613, StickyRay Flat: mean 39.68°, variance 433, RayCursor Flat: mean 45.14°, variance 546, Adaptique Flat: mean 39.19°, variance 365, StickyRay Deep: mean 40.51°, variance 603, RayCursor Deep: mean 35.07°, variance 603, Adaptique Deep: mean 34.43°, variance 500, significant differences: Sparse (** RayCursor and Adaptique), Dense (** StickyRay and RayCursor, StickyRay and Adaptique), Flat (* StickyRay and RayCursor, ** RayCursor and Adaptique), Deep (* StickyRay and RayCursor, StickyRay and Adaptique); 
    (i) Rotational Movement by Technique and Size: StickyRay Small: mean 45.85°, variance 589, RayCursor Small: mean 45.32°, variance 547, Adaptique Small: mean 41.07°, variance 520, StickyRay Large: mean 41.79°, variance 514, RayCursor Large: mean 38.68°, variance 592, Adaptique Large: mean 36.67°, variance 467, ** for small sizes between StickyRay and Adaptique, * RayCursor and Adaptique, * for large sizes between StickyRay and Adaptique; 
    (j) Error Rate by Technique: StickyRay: mean 11.4\%, variance 0.03, RayCursor: mean 6.3\%, variance 0.02, Adaptique: mean 4.9\%, variance 0.01, *** between StickyRay and RayCursor, StickyRay and Adaptique; 
    (k) Error Rate by Technique and Environment: StickyRay Sparse: mean 1.0\%, variance 0.001, RayCursor Sparse: mean 2.8\%, variance 0.004, Adaptique Sparse: mean 1.4\%, variance 0.002, StickyRay Dense: mean 24.0\%, variance 0.03, RayCursor Dense: mean 10.8\%, variance 0.03, Adaptique Dense: mean 9.7\%, variance 0.01, StickyRay Flat: mean 1.7\%, variance 0.003, RayCursor Flat: mean 3.5\%, variance 0.006, Adaptique Flat: mean 2.8\%, variance 0.003, StickyRay Deep: mean 18.8\%, variance 0.03, RayCursor Deep: mean 8.0\%, variance 0.01, Adaptique Deep: mean 5.6\%, variance 0.01, significant differences: Dense (* StickyRay and RayCursor), Deep (* StickyRay and RayCursor); 
    (l) Error Rate by Technique and Size: StickyRay Small: mean 13.0\%, variance 0.03, RayCursor Small: mean 7.1\%, variance 0.02, Adaptique Small: mean 5.2\%, variance 0.01, StickyRay Large: mean 9.7\%, variance 0.02, RayCursor Large: mean 5.4\%, variance 0.01, Adaptique Large: mean 4.5\%, variance 0.01, ** for small sizes between StickyRay and RayCursor, RayCursor and Adaptique, * for large sizes between StickyRay and Adaptique.}
    \label{fig:results}
\end{figure*}

\subsubsection{Selection Time}
We defined the selection time as the time elapsed from the start of the trial to the completion of the selection. We applied a square-root transformation since the distribution of selection time was slightly positively skewed. Significant main effects were observed for \textsc{Technique} \anova{2}{34}{45.27}{<}{.001}{.73}, \textsc{Size} \anova{1}{17}{72.83}{<}{.001}{.81}, and \textsc{Environment} \anova{1.74}{29.50}{374.63}{<}{.001}{.96}. Post hoc analysis of \textsc{Technique} main effect (\autoref{fig:results}a)  revealed that both \system{} and StickyRay were faster than RayCursor (both \ph{<}{.001}).

Additionally, we found no significant three-way interaction. However, we found significant two-way interactions for \textsc{Technique} $\times$ \textsc{Environment} \anova{3.8}{64.60}{15.02}{<}{.001}{.47} and \textsc{Technique} $\times$ \textsc{Size} \anova{2}{34}{16.58}{<}{.001}{.494}. 
Post hoc analysis of \textsc{Technique} $\times$ \textsc{Environment} results (\autoref{fig:results}b) showed that in the \textsc{Sparse} environment, \system{} and StickyRay outperformed RayCursor in speed (both \ph{<}{.001}). In the \textsc{Flat} environment, StickyRay emerged as the significantly fastest technique, followed by \system{}, with RayCursor being the slowest (all \ph{<=}{.003}). In the \textsc{Deep} environment, StickyRay also proved to be faster than RayCursor (\ph{=}{.032}). The techniques did not differ significantly in the \textsc{Dense} environment. For all techniques, users were significantly quickest in the \textsc{Sparse} environment, followed by the \textsc{Flat}, \textsc{Deep}, and \textsc{Dense} environments, the latter resulting in the slowest performance (all \ph{<}{.001}). For the \textsc{Technique} $\times$ \textsc{Size} interaction (\autoref{fig:results}c), both \textsc{Large} and \textsc{Small} targets were selected significantly faster with StickyRay and \system{} compared to RayCursor (all \ph{<}{.001}).

\subsubsection{Movement}
We considered translational and rotational movement for the analysis, defined as the total distance traveled and the angle of rotation of the controller from the start of the trial until the selection was completed. Since both metrics were severely positively skewed, we performed a reciprocal transformation to meet the normality requirement. 

For translational movement, there were significant main effects of \textsc{Technique} \anova{2}{34}{29.48}{<}{.001}{.63}, \textsc{Size} \anova{1}{17}{32.81}{<}{.001}{.66}, and \textsc{Environment} \anova{2.05}{34.87}{53.85}{<}{.001}{.76}. \system{} and RayCursor required significantly less movement compared to StickyRay (both \ph{<}{.001}, \autoref{fig:results}d). Furthermore, no significant three-way interaction was found. Significant two-way interactions were observed for \textsc{Technique} $\times$ \textsc{Environment} \anova{3.68}{62.62}{13.54}{<}{.001}{.46} and \textsc{Technique} $\times$ \textsc{Size}. 
Post hoc analysis of \textsc{Technique} $\times$ \textsc{Environment} (\autoref{fig:results}e) showed that in the \textsc{Sparse}, \textsc{Dense} and \textsc{Deep} environments, \system{} and RayCursor required significantly less movement compared to StickyRay (all \ph{<=}{.008}). While in the \textsc{Flat} environment, we observed that only \system{} required significantly less movement than StickyRay (\ph{=}{.003}). For \textsc{Technique} $\times$ \textsc{Size} interaction (\autoref{fig:results}f), selecting both \textsc{Large} and \textsc{Small} targets with \system{} and RayCursor required less movement than selecting with StickyRay (all \ph{<}{.001}). 

Similarly, for rotational movement, the main effects were also significant for \textsc{Technique} \anova{2}{34}{6.42}{=}{.004}{.27}, \textsc{Size} \anova{1}{17}{54.03}{<}{.001}{.76}, and \textsc{Environment} \anova{1.75}{29.81}{43.69}{<}{.001}{.72}. \textsc{Technique} post hoc analysis (\autoref{fig:results}g) showed that \system{} again had an advantage, requiring significantly less rotational movement than StickyRay overall (\ph{=}{.007}). However, no significant three-way interaction was found. Significant two-way interactions were found for  \textsc{Technique} $\times$ \textsc{Environment} \anova{6}{102}{8.04}{<}{.001}{.32} and \textsc{Technique} $\times$ \textsc{Size} \anova{1.45}{24.67}{6.21}{=}{.012}{.27}. 
Regarding \textsc{Technique} $\times$ \textsc{Environment} (\autoref{fig:results}h), in \textsc{Flat} environments, both \system{} and StickyRay demanded significantly less rotational movement than RayCursor (both \ph{<=}{.036}). In contrast, in \textsc{Dense} and \textsc{Deep} environments, \system{} and RayCursor required significantly less movement than StickyRay (all \ph{<=}{.043}). Additionally, in the \textsc{Sparse} environment, \system{} required significantly less rotational movement than RayCursor (\ph{=}{.006}). 
For StickyRay, the \textsc{Sparse} environment resulted in significantly least rotational movement, followed by \textsc{Deep} and then \textsc{Dense}, with the \textsc{Flat} environment also requiring significantly less movement than \textsc{Dense} (all \ph{<=}{.026}). For RayCursor and \system{}, the \textsc{Sparse} and \textsc{Deep} environments again required significantly less movement, while \textsc{Dense} resulted in the significantly highest movement (all \ph{<=}{.01}). For \textsc{Technique} $\times$ \textsc{Size} interaction (\autoref{fig:results}i), \system{} required significantly less rotational movement than StickyRay regardless of the target size, and required significantly less rotational movement than RayCursor when selecting \textsc{Small} targets (all \ph{<=}{.023}).

\subsubsection{Error Rate}
We defined an error as any trial with at least one missed selection or with a timeout. The error rate was determined by the number of errors divided by the total number of trials within the same condition.  We included all trials in this analysis. We preprocessed the data with an Aligned Rank Transform (ART)~\cite{Wobbrock2011ART} and ART-C preprocessing for post hoc analysis when relevant~\cite{Elkin2021ARTC}.

We found significant main effects for \textsc{Technique} \anova{2}{34}{30.27}{<}{.001}{.64}, \textsc{Size} \anova{1}{17}{16.81}{<}{.001}{.50}, and \textsc{Environment} \anova{3}{51}{109.37}{<}{.001}{.87}. Post hoc analysis (\autoref{fig:results}j) of  \textsc{Technique} revealed that using \system{} and RayCursor resulted in significantly less error rate than using StickyRay (both \ph{<}{.001}).

We found no significant three-way interaction for error rate. However, we observed significant two-way interactions for \textsc{Technique} $\times$ \textsc{Environment} \anova{3.51}{59.72}{13.16}{<}{.001}{.44} and \textsc{Technique} $\times$ \textsc{Size} \anova{2}{34}{5.8}{=}{.007}{.25}. 
Further \textsc{Technique} $\times$ \textsc{Environment} analysis (\autoref{fig:results}k) revealed that in the \textsc{Dense} and \textsc{Deep} environment, StickyRay resulted in significantly higher error rate than RayCursor (\ph{<=}{.023}). While using StickyRay, selecting in a \textsc{Dense} and \textsc{Deep} environment results in significantly more errors than selecting in a \textsc{Sparse} and \textsc{Flat} environment (all \ph{<}{.001}). Analyzing the \textsc{Technique} $\times$ \textsc{Size} interaction (\autoref{fig:results}l), we found that when selecting \textsc{Small} targets, \system{} and RayCursor resulted in significantly fewer errors than using StickyRay (all \ph{<=}{.003}). When selecting \textsc{Large} targets, \system{} resulted in a significantly lower error rate than StickyRay (\ph{=}{.001}).

\definecolor{darkgreen}{rgb}{0.0353, 0.9216, 0.2824}
\newcommand{\tcheck}{{\color{darkgreen} \checkmark}}
\newcommand{\tcross}{{\color{red} $\times$}}

\begin{table}[t]
\caption{Overview of selection technique comparison across performance metrics. \tcheck{} indicates the techniques that were found most optimal for the given metric (as determined by statistical main effects). \tcross{} indicates less performant techniques. Results showed that \system{} was among the most optimal techniques for all metrics.}
\centering
\begin{tabular}{l|ccc}
\toprule
Metric & StickyRay & RayCursor & \system{}\\
\midrule
Selection Time & \tcheck{} & \tcross{} & \tcheck{}\\
Trans. Movement & \tcross{}& \tcheck{}& \tcheck{}\\
Rot. Movement &\tcross{} & \tcross{}& \tcheck{}\\
Error rate & \tcross{}& \tcheck{}& \tcheck{}\\

\bottomrule
\end{tabular}
\label{table:optimal-selection-techniques}
\Description{
This table provides an overview of the comparison of selection techniques across four performance metrics: Selection Time, Translational Movement, Rotational Movement, and Error Rate. The techniques compared are StickyRay, RayCursor, and Adaptique. For each metric:
- For Selection Time: StickyRay and Adaptique were among the most optimal techniques, while RayCursor was less performant.
- For Translational Movement: RayCursor and Adaptique were most optimal, while StickyRay was less performant.
- For Rotational Movement: Only Adaptique was among the most optimal techniques, with StickyRay and RayCursor being less performant.
- For Error Rate: RayCursor and Adaptique were among the most optimal techniques, while StickyRay was less performant.
Overall, the results showed that Adaptique was consistently among the most optimal techniques for all metrics.
}
\end{table}

\subsubsection{Summary of Quantitative Results}
Our quantitative study results showed that StickyRay and RayCursor exhibit different performance advantages under different conditions and metrics. Overall, StickyRay is faster than RayCursor in all environments except for \textsc{Dense} environment, while RayCursor is more precise and requires less translational movement, especially in \textsc{Dense} and \textsc{Deep} environments.
In contrast, \system{} consistently achieved optimal performance for all metrics in different contexts, as shown in~\autoref{table:optimal-selection-techniques}. Although single techniques occasionally performed as well as \system{} in specific metrics, they exhibited performance degradation in other metrics or a particular environment. For example, although StickyRay performed as well as \system{} in selection time, it required more movement and caused more errors, especially in \textsc{Dense} and \textsc{Deep} environments. Similarly, RayCursor performed comparably to \system{} in terms of movement and accuracy but required significantly more selection time, and even worse in \textsc{Sparse} and \textsc{Deep} environment. \system{} performed comprehensively well in all metrics, indicating that our system effectively balanced between different objectives and technique trade-offs.

\subsubsection{Questionnaire Results and Preferences}
Friedman tests on usability ratings showed significant results in perceived Precision ($\chi^{\scriptscriptstyle 2}(2)$=$7.26$, $p$$<$$.05$), Difficulty ($\chi^{\scriptscriptstyle 2}(2)$=$7.00$, $p$$<$$.05$), and Confidence ($\chi^{\scriptscriptstyle 2}(2)$=$6.26$, $p$$<$$.05$). However, Wilcoxon post hoc tests with Bonferroni correction did not show any significant differences. 

The majority of the participants (ten) preferred \system{} over the other techniques, while four chose StickyRay and four chose RayCursor. \system{} was considered ``fast''(P6), ``precise'' (P13), ``easy to navigate'' (P14), and "convenient" (P3), and combined the advantages of StickyRay and RayCursor, offering the most suitable technique for the environment (5 out of 18). P10 mentioned ``both StickyRay and RayCursor are convenient in different situations. [..] So being able to switch to the other based on situations is preferred''.  Meanwhile, although StickyRay was ``intuitive''(P1) and ``easy to use''(P12), the technique was less ``precise'' (P14) and ``forces users to move more'' (P8) in complex environments with target occlusion (expressed by 11 out of 18 participants). 
Regarding RayCursor, although participants liked its ``full control'' (P5) of cursor depth in cluttered environments, it was considered ``harder'' (P1) and more ``tiring'' (P2)  due to the ``additional control required'' (P8), especially when there were fewer objects (9 out of 18 participants). 
Some participants did not prefer \system{} because it incorporated the technique they did not like, or due to ``delayed'' (P5), or ``distracting'' (P11) switching. For example, P1 said ``\system{} is uncomfortable because it incorporated RayCursor''.

Most of the participants gave positive feedback on the switching (14 out of 18) due to its ``consistency'' (P8) and ``accuracy'' (P16) in technique selection, and ``clear'' (P14) feedback. Participants liked that \system{} only switched techniques when needed and not in the middle of a selection. P8 mentioned that ``the switching is pretty handy and intelligent, selecting the most efficient mode almost all the time. The mode does not vary constantly and is consistent enough for the user to get used to the selection.'' 
Overall, visual, audio, and tactile feedback helped the participants in ``notifying the technique change'' (P4), and ``improving the experience'' (P2). 

\subsubsection{Technique Switching Analysis}

To validate \system{}'s switching proficiency, we analyzed its switching behavior across all trials. One or more switches occurred in 53. 65\% of the trials. In other trials, the current technique was considered optimal, indicating that switching occurs only when \system{} deems it necessary. For technique switching trials, we observed that the number of switches was low, with a mean of 1.1 among the trials involving switching. These results indicate \system{}'s stability in selecting technique, as too sensitive switching may be confusing for users. \system{} also showed strong consistency in selecting the technique according to the current context. Specifically, 99.6\% of \textsc{Deep} environment and 100\% of \textsc{Dense} environment trials concluded with RayCursor, while 98.3\% of \textsc{Flat} and 96.2\% of \textsc{Sparse} environment trials concluded with StickyRay. 

In terms of timing, we found that the first switch typically occurred early in the interaction, on average at 0.37 seconds (variance = 0.08) after the trial began, compared to an average total task duration of 1.67 seconds. To further investigate the temporal relationship between switching and user movement, we compared user movement with the timing of switches. In most cases, the switch occurred before the onset of their pointing action. These results suggest that \system{} is able to proactively switch techniques before users initiate a pointing movement, thus minimizing disruption and cognitive load. However, we also observed a small number of cases where switching occurred during or after the initial ballistic movement, possibly due to occasional delays in calculations or users pointed beyond the trial area. In these instances, users were usually able to complete the selection without initiating another pointing movement. 
However, in rare cases (8.5\% of the trials in which a switch occurred), a second high-speed movement occurred, likely as a corrective adjustment following the delayed switch.

Overall, these findings indicate that \system{} switches techniques efficiently, consistently, and early enough to assist users without imposing cognitive overhead or disruption to movement.
%

\section{Discussion}
We introduced ~\system{} to address the selection challenges inherent in dynamic virtual environments. \system{} adapts the selection technique according to a wide range of environments and user states based on a computational approach of extracting contextual information that effectively captures scenarios where users perceive objects as overlapping, too small, arranged differently, etc. \system{} also considers different aspects of performance built from established selection models and balances these factors to align with the design needs.

Our results underscore the need for adaptivity, as using the same technique in various scenarios can introduce trade-offs that lead to difficulty and negatively impact performance and user experience. For example, although RayCursor is effective in precisely selecting objects, especially in \textsc{Dense} and \textsc{Deep} environments, it is generally slower due to its complexity in selecting objects, especially in \textsc{Sparse} and \textsc{Flat} environments. In contrast, \system{} consistently achieved optimal results in terms of selection time, movement, and error rate in different contexts (\autoref{table:optimal-selection-techniques}). Although single techniques can perform as well as \system{} in specific metrics, they typically exhibit performance degradation in other metrics or particular environments. These findings suggest that our method successfully identifies and applies the most suitable technique for each scenario to balance objectives and technique trade-offs.

Our application showcases \system{}'s utility and applicability in a dynamic and practical setting. ~\system{} can automatically switch the selection tool to a more suitable one when the task becomes more difficult to complete with the current tool. For example, selecting book layouts on the bookshelf is easy with normal RayCasting, while selecting books stacked on the table is difficult because they can occlude one another and require extra precision. Therefore, when users point toward the stack of books, ~\system{} smoothly and proactively switches to RayCursor, ensuring smooth transitions and consistent tool usage in the new context. This adaptive behavior is driven by the interaction space spreading out from the pointing direction, which gradually captures the context of the user's attention. In addition, the thresholding mechanism ensures consistent improvement across frames before confirming a switch, preventing users from experiencing inconsistent switching.

\subsection{Contextual Information Extraction}
Adaptique effectively extracts fundamental contextual information as the basis for adaptation. In contrast to previous work that only extracts density information~\cite{cashion2013Optimal}, we further parameterize target relationships, such as occlusion, to model objectives more accurately. This consideration is crucial because the performance depends on an object's effective size rather than its dimensions.
Still, although our current approach considers occlusion relative to the controller's perspective as this directly influences performance, this may not align with how users perceive occlusion. For example, an object that appears occluded from the user's sight might not be occluded according to the controller's viewpoint. 
This discrepancy can create a mismatch of the optimal technique between user expectations and the model's behavior, degrading the user experience more than the performance benefits justify. Future work could explore integrating perceived contextual information to enhance user intuitiveness.

Additionally, although \system{} is currently implemented in VR, it is intended for future extension into mixed reality environments, where physical interactable objects are rigid and the surroundings are more dynamic and unpredictable. However, there remain limitations to the tracking of the 3D positions and shapes of real-world objects. We believe that future improvements in object tracking technologies will overcome these limitations~\cite{kirillov2023Segment, ravi2024sam2segmentimages}, allowing more use cases to benefit from ~\system{}.

\subsection{Optimization Objectives}
Adaptique integrates multiple objectives that balance performance (e.g., speed, accuracy) and usability (e.g., comfort, familiarity). This formulation, combined with contextual modeling, generalizes well to diverse XR scenarios.
Building on this foundation, \system{} can be further expanded to include more contextual information and objectives. For example, including the moving speed of targets or users~\cite{hasan2011Comet, li20182DBayesPointer, Manakhov2024Gazeonthego} can be beneficial in scenarios such as public transportation, interactions with moving targets like people and animals, and gaming. Additionally, the contextual information could be expanded to account for factors accumulated over time, such as cumulative fatigue~\cite{johns2023Pareto, evangelistabelo2021XRgonomics} and workload~\cite{lindlbauer2019ContextAware} during prolonged interactions. More objectives can also be considered to accommodate different design requirements, such as social acceptance~\cite{eghbali2019Social, williamson2019PlaneVR, situationadapt2024li}, engagement~\cite{pohl2013Focused}, sensor error~\cite{sidenmark2020Outline}, available range of motion~\cite{williamson2019PlaneVR, cheng2023InteractionAdapt}, and so on.

To enable adaptation in real-time, our objectives include certain assumptions and simplifications. For example, the body parameters for calculating comfort scores rely on average human data instead of individual data. While this simplification might introduce deviation from using individual parameters, the error affects all techniques equally and, therefore, has little impact on our adaptation.
Alternatively, in the future users could also input their specific values for more precise estimation.
In addition, we assumed users would select the target with a specific movement trajectory due to the challenges of predicting user motions. Though users might not perform the selection in this manner, this approach captures the general trends of exertion and fatigue and responsively adapts to users' posture changes. Future work could include a movement prediction model based on velocity profiles~\cite{henrikson2020HeadCoupled}.
Our approach to assigning familiarity scores through pilot testing proved effective and practical in our study, but it may have limitations in scalability and subjectivity, as further pilot testing is required for different sets of techniques. Further research could develop a more systematic approach to assigning familiarity scores, such as inferring values from user preferences collected through voting on various benchmark scenes. Another direction could be training machine learning models to predict the complexity of selection techniques using behavioral data, such as cognitive load metrics~\cite{octavia2011Squeeze}.

\subsection{Selection Techniques}
Currently, we considered raycasting-based selection techniques as candidates that address various scenarios effectively. 
Future explorations could include more diverse pointing-based techniques, such as those with different modalities~\cite{sidenmark2022Weighted, Chen2023GazeRayCursor}. This would open up a vast design space to integrate complementary techniques.
For example, switching the technique to gaze when the arm is tired or occupied can be useful in prolonged use. However, modifications to the current objective score calculations would be necessary to ensure these scores remain comparable across different modalities.

To reduce confusion and required training, we limited the number of techniques to one per selection scenario. Although having more selection techniques as candidates increases the granularity of optimal choices, we suggest it could lead to excessive switchings that distract the user. While this limitation reduces distraction, future exploration could provide alternative techniques for the same selection scenario to allow users to customize their set of techniques. This flexibility would address feedback from study participants who did not prefer \system{} as it included techniques they disliked.

Additionally, we encountered challenges in modeling advanced selection techniques. For example, RayCursor, which is controlled by a combination of ray movement and swiping on the trackpad, makes the user interaction pattern unpredictable. Thus, we simplified the model by considering only the movement of the ray and assuming the cursor remains at the correct depth. This simplification might overestimate the technique's performance as the swiping effort is not modeled, and we assume the calibration effect of the familiarity score mitigated this deviation. Future research could explore more sophisticated models for advanced techniques~\cite{Adrian2017Fitts} or introduce a calibration method for the model that could not fully describe a technique. Another possible approach could be to employ data-driven methods to better understand the relationship between human behaviors and performance metrics.

\subsection{Further Adaptation Considerations}
Our qualitative result showed that participants were satisfied with the switching due to its consistency and accuracy in technique selection. 
Similarly, our quantitative results showed no significant performance degradation from potential switching or distraction side effects, as Adaptique consistently performed among the best across conditions.
This outcome was achieved by optimizing system responsiveness and adjusting sensitivity through a window threshold technique to minimize delays and unnecessary switching. To further enhance switching, we could combine the current system with an intention prediction model to decide the optimal switching timing~\cite{yu2022Optimizing}. 

However, even though we did not identify the significant cost of switching, there might still exist subtle side effects that our current evaluation could not adequately capture. This raises the possibility that Adaptique, under optimal conditions, might outperform individual techniques, rather than merely showing comparable performance to the most effective one. To further investigate this, future studies could focus on quantifying switching costs by measuring human response time and cognitive load. These measures could also be analyzed in relation to the complexity of the techniques being switched or the degree of disparity between them~\cite{scarr2011Dips}.

\section{Conclusion}
We presented \system{}, an online multi-objective model that adaptively switches to the most optimal VR selection technique based on user context and environment combined with human performance objectives. The results show that \system{} can significantly improve selection time, movement, and error rate against the use of singular techniques. In addition, a majority of participants preferred \system{} who expressed a positive sentiment for switching techniques when exposed to various environments. In sum, \system{} shows that it is beneficial to switch between techniques to gain the most performance across multiple environments. Furthermore, considering multiple objectives is important to reflect the trade-offs between different techniques. Our work opens up further research on additional selection objectives, techniques, and modalities to accurately model and adapt to interactions commonly needed in our daily lives.




\bibliographystyle{ACM-Reference-Format}
\bibliography{ref}


\end{document}